\documentclass[times,twocolumn]{aastex62}
\pdfoutput=1 
\usepackage{amsmath,amstext}
\usepackage[T1]{fontenc}
\usepackage[figure,figure*]{hypcap}

\usepackage{bm}
\usepackage{graphicx}	
\usepackage{amsmath}	
\usepackage{amssymb}	
\usepackage{longtable}
\usepackage{subfigure}
\usepackage{rotating}
\usepackage{float}
\usepackage{natbib}
\usepackage{xspace}
\usepackage{epstopdf}
\usepackage[]{url}
\usepackage{enumitem}

\newcommand{\Msun}{M$_{\odot}$\xspace}
\newcommand{\Mvir}{M$_{\rm vir}$\xspace}

\newcommand{\Rvir}{R$_{\rm vir}$\xspace}

\newcommand{\kms}{km s$^{-1}$\xspace}
\newcommand{\kpckms}{kpc km s$^{-1}$\xspace}

\shorttitle{MW Mass Estimates Using Classical Satellites}
\shortauthors{Patel et al.}

\begin{document}

\title{Estimating the Mass of the Milky Way Using the Ensemble of Classical Satellite Galaxies}

\author{Ekta Patel}
\author{Gurtina Besla}
\affiliation{Department of Astronomy, University of Arizona, 933 North Cherry Avenue, Tucson, AZ 85721, USA; ektapatel@email.arizona.edu}

\author{Kaisey Mandel}
\affiliation{Institute of Astronomy and Kavli Institute for Cosmology, Madingley Road, Cambridge, CB3 0HA, UK}
\affiliation{Statistical Laboratory, DPMMS, University of Cambridge, Wilberforce Road, Cambridge, CB3 0WB, UK}

\author{Sangmo Tony Sohn}
\affiliation{Space Telescope Science Institute, 3700 San Martin Drive, Baltimore, MD 21218, USA}

\begin{abstract}
High precision proper motion (PM) measurements are available for approximately 20\% of all known dwarf satellite galaxies of the Milky Way (MW). Here we extend the Bayesian framework of \citet{patel17b} to include all MW satellites with measured 6D phase space information and apply it with the Illustris-Dark simulation to constrain the MW's mass. Using the properties of each MW satellite individually, we find that the scatter among mass estimates is reduced when the magnitude of specific orbital angular momentum (\textit{j}) is adopted rather than their combined instantaneous positions and velocities. We also find that high \textit{j} satellites (i.e. Leo II) constrain the upper limits for the MW's mass and low \textit{j} satellites rather than the highest speed satellites (i.e. Leo I and LMC), set the lower mass limits. When \textit{j} of all classical satellites is used to simultaneously estimate the MW's mass, we conclude the halo mass is $0.85^{+0.23}_{-0.26}\times 10^{12}$ \Msun (including Sagittarius dSph) and $0.96^{+0.29}_{-0.28} \times 10^{12}$ \Msun (excluding Sagittarius dSph), cautioning that low \textit{j} satellites on decaying orbits like Sagittarius dSph may bias the distribution. These estimates markedly reduce the current factor of two spread in the mass range of the MW. We also find a well-defined relationship between host halo mass and satellite \textit{j} distribution, which yields the prediction that upcoming PMs for ultra-faint dwarfs should reveal \textit{j} within $ 5\times10^3 - 10^4$ \kpckms. This is a promising method to significantly constrain the cosmologically expected mass range for the MW and eventually M31 as more satellite PMs become available.
\end{abstract}


\section{Introduction}
Satellite galaxies around the Milky Way (MW) are often used to study the structure and assembly history of the Galaxy's dark matter and stellar halo. In particular, the kinematics of halo tracers (satellites, globular clusters and stellar streams) have been used to constrain the Galaxy's gravitational potential and total mass.  Many efforts have been made to constrain the current factor of two spread in the total mass range of the MW, but a high precision estimate has yet to be made. 

Leo I, as one of the highest speed MW satellites, is often used to place lower limits on the mass of the MW.  However, because its relative velocity hovers around the MW's escape speed at its current separation of $\sim$260 kpc, any MW mass constraint requires the assumption that Leo I is bound. \citet[][hereafter BK13]{bk13} illustrate that unbound satellites within the virial radius are rare. By assuming that Leo I is bound at its current position and velocity, BK13 used the Aquarius cosmological zoom simulations \citep{springel08a} to infer that the virial mass of the MW must be > $10^{12}$ \Msun, with a median mass of $1.6\times10^{12}$ \Msun. Other studies relying on the boundedness of Leo I, such as \citet{liwhite08}, used a radial timing argument analysis similar to that of \citet{zaritsky89} to determine that the MW's mass is $2.43\times10^{12}$ \Msun with a lower limit of $0.8\times 10^{12}$ \Msun. Non-radial timing argument studies predict a MW mass as high as $> 3\times10^{12}$ \Msun if Leo I is bound \citep{sohn13}.  

If Leo I is unbound, however, these methods would likely overestimate the true mass of the MW. The case of Leo I strongly motivates a study to calibrate the impact of using a single satellite to constrain the mass of its host. We argue in this study that a single satellite can result in a significantly biased mass estimate, especially when satellites are on extreme orbits \citep[see also][]{sales07, patel17b}. Instead, we have developed a novel method of estimating the mass of the MW using an ensemble of observed satellites and analogs in cosmological simulations. 

The HST Proper Motion (HSTPROMO) collaboration \citep[e.g.]{sohn17, k13, sohn13, massari13, k06a, k06b} and other authors \citep{piatek16, piatek08, piatek07, piatek06, piatek05, piatek03, piatek02, walker08, scholz94} have now measured the proper motions of several low mass dwarf galaxies in the MW's halo. With the combined 6D phase space information derived from the proper motions for the classical MW satellites (LMC, SMC, Carina, Draco, Fornax, Sculptor, Leo I, Leo II, Ursa Minor, Sextans, and the Sagittarius dSph), we can now use the 3D dynamics of a population of halo tracers to further our understanding of the MW's dark matter halo and its global properties. In this work, we utilize the magnitude of specific orbital angular momentum of each satellite, rather than their instantaneous position and velocity, as motivated in \citet[][hereafter P17B]{patel17b}. An important benefit is that we make no assumption of whether a satellite is bound, simply whether it currently resides within the virial radius of the MW. 

As the two most massive MW satellites, the Magellanic Clouds (MCs) are also often used to characterize properties of the MW, specifically via analogs selected from cosmological simulations. Both \citet{bk11} and \citet{busha11} have used the instantaneous characteristics of the MCs, such as their current position and velocity, to make predictions for the MW's mass in a frequentist and Bayesian fashion, respectively \citep[see also][]{gonzalez13}. Using the \citet{k13} proper motions, the Bayesian posterior mean mass estimate for the MW using the Large Magellanic Cloud (LMC) alone is $1.70 \times 10^{12}$ \Msun (P17B) and approximately $1.15\times 10^{12}$ \Msun using both MCs \citep{gonzalez13}. 

While the MCs are well-studied members of the MW's halo, their current properties including the spatial proximity of the MCs to the MW  ($\sim$50 kpc), to each other ($\sim$23 kpc), and their unique orbital configuration (just past pericenter), make them rare in a cosmological context \citep{b07, besla12, bk11}. Less than 5\% of simulated MW mass halos host two massive satellites that have made a recent first pericentric approach as close as 50 kpc. We showed in P17B that MW mass estimates are highly susceptible to the orbital phase of the LMC and thus conclusions based on the current properties of only the MCs should be taken with caution. As a result, we must turn to other satellite properties to estimate the mass of the MW in an unbiased fashion. 

The \emph{momentum method} developed in P17B relies on orbital angular momentum and thus affords a larger simulated data set from which MW mass estimates can be formed since it does not limit satellite analogs to a narrow range of position and velocity combinations as in the \emph{instantaneous method}. For the MW-LMC system, a nearly ten-fold increase in the number of statistically significant satellite analogs from the dark-matter-only Illustris simulation (or Illustris-1-Dark) enables us to combine inferences for the MW's mass even for rare systems (i.e. unique orbital configurations), thus providing a powerful method moving forward as more high precision data becomes available for halo tracers.

The impact of only one satellite on the mass estimate of the MW is clearly evident with regards to Leo I. This motivates the need to use several satellites simultaneously to form a MW mass estimate. Here, we extend the P17B method to all classical low mass MW satellites (satellites less massive than the MCs) to constrain the mass of the MW. This will test whether satellites like Leo I are still outliers in these MW mass estimation techniques or if other satellites will become more critical players. For example, recent studies by \citet{gibbons14, belokurov14} have suggested a lower mass bound for the MW based on the properties of the Sagittarius dSph stellar stream, indicating a lower limit of about $7\times10^{11}$ \Msun. With this statistical framework, such assertions can be tested in a cosmological context for the first time. 

In Section \ref{sec:simandobsprops}, we compile properties of the nine low mass satellites used in this study and describe the details of the Illustris dark-matter-only cosmological simulation. Section \ref{sec:statmethods} provides the details of the Bayesian framework that has been extended from P17B to accommodate a population of lower mass satellites. The results of these methods are presented in Section \ref{sec:results}. Section \ref{sec:discussion} contains a discussion of the global trend between the distribution of satellite specific orbital angular momenta and host halo mass. Finally, Section \ref{sec:conclusion} summarizes our results.

\section{Simulation and Observed Satellite Properties}
\label{sec:simandobsprops}

In P17B, we combined the properties of the most massive satellite galaxies in the halos of the MW and M31(the LMC and M33) with a state-of-the-art cosmological simulation in a Bayesian statistical framework to infer the masses of their host halos. In this work, we extend our previous analysis to lower mass satellites and additionally implement a method that uses multiple satellites simultaneously to refine MW mass estimates. This section describes the observational data that is currently available for the low mass MW satellites considered and the specifications of the Illustris-1-Dark simulation, which we will use to select low mass satellite analogs.

\subsection{Observed Properties of Nine Low Mass MW Satellites}
\begin{table*}
\begin{center}
\caption{Observational MW Satellite Data}
\begin{tabular}{ccccccccc} \hline \hline
& $\rm r^{obs}$ & $1\sigma$ & $\rm v_{tot}^{obs}$ & $1\sigma$ & $\rm j^{obs}$& $1\sigma$ & References \\ 
& [kpc] & [kpc]  & [\kms] & [\kms] & [kpc \kms] & [kpc \kms] & Dist. + $\rm v_{rad}$ & PM \\ \hline
Leo II & 236 & 14 & 138 & 42 & 32,105 & 9,998 & 1& 4\\ 
Fornax & 149 & 12 & 193 & 45 & 28,471 & 8,712 &1& 5\\ 
Leo I & 258 & 15 & 202 & 19 & 27,918 & 9,146 & 1 & 3\\
Sculptor & 86 & 6 & 213 & 10 &17,232 & 1,267 & 1 & 2 \\ 
Sextans & 89 & 4 & 206 & 22 & 17,097& 2319 & 1 & 8\\ 
LMC & 50 & 5 & 321 & 24 & 15,688 & 1,788 & 1,10 & 10\\ 
Ursa Minor & 78 & 3 & 173 & 49 & 12,545 & 4,375 & 1 & 7\\ 
Draco & 76 & 6 & 183 & 4 & 12,241 & 1,229 & 1 & 2\\ 
Carina & 107 & 6 & 97 & 41 & 10,322 & 4,532 & 1 & 6\\ 
Sagittarius dSph & 18 & 2 & 320 & 23 & 5,249 & 1,000 & 1 & 9 \\ \hline \hline
\end{tabular}
\end{center}
\tablecomments{Observational data (${\bm d}$) for the LMC and the nine MW dSph galaxies with proper motions (PMs). The satellites are listed in order of descending $\rm j^{obs}$. Data for the distance, radial velocity, and PMs of each satellite were taken from the following references.}
\tablerefs{(1) \citet[][and references therein]{mcconnachie12}, (2) \citet{sohn17}, (3) \citet{sohn13}, (4) \citet{piatek16}, (5) \citet{piatek07}, (6) \citet{piatek03}, (7) \citet{piatek05}, (8) \citet{casetti17}, (9) \citet{sohn15}, (10) \citet{k13}. }
\label{table:bayesparams}
\end{table*}

To date, the proper motions of the nine brightest MW dwarf satellite galaxies (after the MCs) have been measured with high astrometric precision in the last decade. In principle, any halo tracer (e.g., satellite galaxies, globular clusters) with proper motion information can be used in this type of statistical analysis to estimate the mass of its host if a sufficiently large set of simulated analogs are also available. 

The classical\footnote{Classical in this case refers to those satellites that were known prior to the Sloan Digital Sky Survey. These satellites are also the brightest (and most massive) MW satellites aside from the Magellanic Clouds (MCs).} low mass satellites considered in this work include: Carina, Draco, Fornax, Sculptor, Leo I, Leo II, Ursa Minor, Sextans, and the Sagittarius dSph. The stellar masses of these dwarf spheroidal (dSph) galaxies are between $10^5-10^7$ \Msun \citep[see][]{mcconnachie12} and their masses at infall are predicted to be $10^8-10^{10}$ \Msun from cosmological expectations \citep{bk12, moster13}. For reference, the LMC's stellar mass is about $3\times10^9$ \Msun at present and its total mass at infall could be as high as a few times $10^{11}$ \Msun \citep{kim98, besla10, wang06}. 

The observed data (${\bm d}$) for each low mass satellite and the LMC are provided in Table~\ref{table:bayesparams}. These data consist of: i.) the observed position relative to the MW ($ r^{\rm obs}$), ii.) the total velocity relative to the MW ($ v^{\rm obs}_{\rm tot}$), and iii.) the magnitude of the specific orbital angular momenta about the MW ($ j^{\rm obs}$), where $j = |{\bf r} \times {\bf v}|$. The mean values of these quantities and the uncertainties associated with them have been calculated from a set of 10,000 Monte Carlo samples drawn from the $4\sigma$ error space of distance, radial velocity, and proper motion of each satellite. For the low mass satellites, all adopted distances and radial velocities are taken from the compilation presented in \citet{mcconnachie12} and the references therein. The proper motions of the galaxies come from a variety of groups and programs (most measurements come from the {\em Hubble Space Telescope}), as indicated in the final column of Table~\ref{table:bayesparams}. In this analysis we do not use the properties of the SMC explicitly due to the low frequency of LMC and SMC analogs in cosmological simulations \citep[see][]{bk11}, but SMC analogs may exist about the halos included in the prior sample (see Section \ref{subsec:prior}).

\subsection{The Illustris-Dark Simulation}

To choose a broad set of MW halo analogs, we use the halo catalogs from the publicly available Illustris Project, a suite of N-body+hydrodynamic simulations run with the \texttt{AREPO} code \citep{vogelsberger14a, vogelsberger14b, genel14, nelson15}. For this analysis, we use only the Illustris-1-Dark (hereafter Illustris-Dark) simulation, which spans a cosmological volume of (106.5 Mpc)$^3$ and follows the evolution of 1820$^3$ dark matter particles from redshift $z=127$ to $z=0$. Each dark matter particle has a mass of $m_{\rm DM}=7.5\times10^{6}$ \Msun. Hydrodynamical processes are not included in the main body of this analysis because Illustris-Dark affords a larger set of cosmological analogs. However, we have included a comparison to the Illustris-1 hydrodynamics simulation in Appendix \ref{sec:appendix2}. All Illustris simulations use the {\em WMAP-9} cosmological parameters \citep[see][]{hinshaw13}.

Substructures in each of the 136 snapshots of the Illustris-Dark simulation are identified using the \texttt{SUBFIND} algorithm \citep{springel01, dolag09}. \texttt{SUBFIND} is a halo-finding routine that first identifies halos using a friends-of-friends (FoF) method and then finds substructures within each identified halo that are overdense, gravitationally bound regions. Typically, each FoF group contains a massive, central subhalo that contains most of the loosely bound material in the halo. A selection of these centrals will act as MW analogs in this work. The Illustris-Dark merger trees were created using the \texttt{SUBLINK} code \citep{rg15}. All relative positions for halos and subhalos are corrected for the box edges. 

In this work, we will refer to the virial mass and radius of FoF groups in our sample. While the virial mass is the mass of all substructures in a FoF group, it is approximately equivalent to the mass of the primary, central subhalo in each halo. Virial mass is defined as the mass enclosed within the virial radius -- the radius at which the average density within that radius reaches an overdensity of $\rm \Delta_{vir}$ in a spherical`top-hat' perturbation model. For a $10^{12}$ \Msun halo, this corresponds to a virial radius of about 260 kpc. The $\rm \Delta_{vir}$ factor depends directly on the cosmological parameters \citep[see][]{brynorman98}. The Illustris-Dark cosmology yields $\rm \Delta_{vir}$ = 357 (or $\rm \Delta_{vir} \times \Omega_m$ = 97.3). The virial mass and radius are taken directly from the Illustris-Dark group catalogs. A variety of other mass definitions, such as those based on the splashback radius or $R_{200}$, could also be used in this analysis.

\section{Statistical Methods}
\label{sec:statmethods}

In this section, we identify the subset of host halos from P17B's prior that have exactly one massive satellite analog and at least one satisfactory low mass subhalo analogous to the classical satellites. Note that while the MW actually hosts several low mass satellites, we only require a minimum of one low mass satellite analog per prior sample host halo due to the simulation's resolution limits. We also outline the modified selection criteria ($\bf{C'}$) for the prior sample, likelihood functions tailored for low mass satellites, and the statistical approximation used to infer the mass of the MW using the ensemble of low mass satellites simultaneously.

\subsection{Prior Samples}
\label{subsec:prior}

\begin{figure*}
\begin{center}
\includegraphics[scale=0.45, trim=0mm 0.75cm 0mm 0mm]{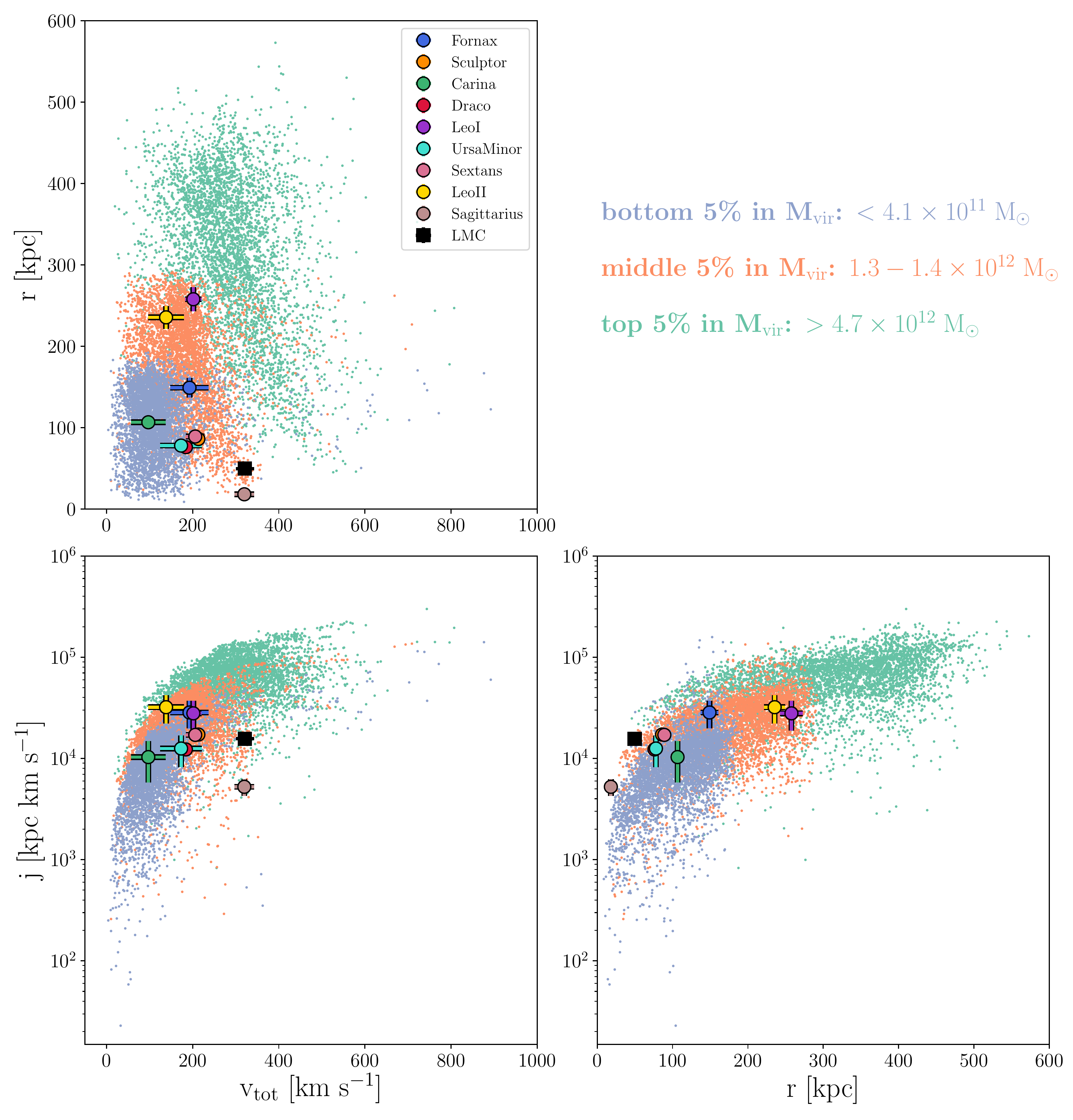}
\caption{For the low mass satellite analogs selected to be in Prior 2, the distribution of satellite subhalo properties (${\bm x}$) are shown for each pair of satellite parameters. Only those properties for low mass satellite analogs residing in the bottom 5\% ($< 1.4 \times 10^{11}$ \Msun), middle 5\% ($1.3-1.4 \times 10^{12}$ \Msun) and top 5\% ($> 4.7 \times 10^{12}$ \Msun) of the host halo mass (\Mvir) distribution are shown to illustrate the spread in satellite subhalo properties as a function of host halo mass. The colored points with error bars denote the observed properties (${\bm d}$) for all satellites listed in Table \ref{table:bayesparams}. The overlap between the observed satellite properties and the properties of the prior sample indicate that our prior selection criteria are chosen appropriately.
\label{fig:priorscatterothersats}}
\end{center}
\end{figure*}

\subsubsection{Massive Satellite Analogs}
\label{subsubsec:massivesats}
While this work focuses on the low mass MW satellites, it is important to include the presence of a massive satellite analog (weighing approximately 10\% of their host's total mass) since they could alter the gravitational potential of their host halos and subsequently affect the kinematics of other nearby satellites (see Section \ref{sec:discussion}). The existence of a massive satellite analog also ensures that MW analogs that have similar mass assembly histories are chosen. We also note that the subhalo abundances around host halos with and without a massive satellite analog differ, which we discuss in detail in Appendix \ref{sec:appendix1}.

To build a prior sample, we first require that exactly one massive satellite analog satisfying the following selection criteria ($\bf{C}$) from P17B exists in all halos from the final 20 snapshots ($z\approx 0.26$ to $z=0$) of the Illustris-Dark simulation. We consider only the final 20 snapshots to be consistent with P17B and previous work \citep[e.g.,][]{busha11,gonzalez13}.

\begin{itemize}
\item[] C$_1$: A subhalo is considered a massive satellite analog only if \indent\indent $v_{\rm max} >$ 70 \kms.
\item[] C$_2$: The massive satellite analog must reside within its host's \indent\indent virial radius (\Rvir) at $z\approx0$.
\item[] C$_3$: The massive satellite analog must have a minimal subhalo \indent\indent  mass of $10^{10}$ \Msun at $z\approx0$. 
\end{itemize}

Note that subhalo mass is provided in units of \Msun $h^{-1}$, where $h=0.704$, in the Illustris-Dark halo catalogs and is used as given. $v_{\rm max}$ is the maximum circular velocity of a subhalo. These selection criteria return a total of 19,653 host halos, each with a companion massive satellite analog. This data constitutes the prior sample used in P17B.

\subsubsection{Low Mass Satellite Analogs}
\label{subsubsec:lowmasssats}
Analogs of the low mass classical MW satellites must belong to the first 15 subhalos in each FoF group, which are ranked by decreasing subhalo mass. This ensures that systems with multiple {\em massive} satellite analogs (see previous section) are omitted. Ideally, the prior sample would contain only those halos that host a massive satellite analog in addition to about 10 low mass satellite analogs. However, simulations with large volumes and sufficient resolution are currently unable to resolve a statistically significant number of halos that represent true analogs of the MW in this way, so these constraints are relaxed. Small scale cosmological volumes with higher mass resolution and perhaps finer redshift spacing are likely the most ideal simulations to build such a prior sample in the future.

We require that all low mass satellite analogs have $v_{\rm max} < 45$ \kms to ensure that we are truly selecting just the low mass subhalos (i.e. the MCs have $v_{\rm max} > 50$ \kms). To avoid the `Too Big to Fail' (TBTF) problem \citep{bk11b}, only an upper $v_{\rm max}$ bound is used in these criteria. TBTF is a discrepancy that arises in dissipation-less $\Lambda$CDM simulations where subhalos with masses analogous to the classical MW dwarfs fail to host bright satellites. This discrepancy leads to a mismatch between simulated $v_{\rm max}$ values and their observed counterparts such that the observed values are much lower than that of their simulated analogs. 

Our goal in this work is to choose dynamical analogs, not analogs based purely on mass or energetics. Since dynamical friction plays an insignificant role in the orbital evolution of these low mass satellites, the kinematics (positions, velocities, specific angular momenta) are not expected to decrease drastically between infall and today. The selection criteria for low mass satellite analogs (denoted by ${\bf C'}$) are summarized below. 

\begin{itemize}
\item[] C$'_1$: A subhalo is considered a low mass satellite analog only \indent\indent if $ v_{\rm max} <$ 45 \kms.
\item[] C$'_2$: The satellite analog must reside within its host's virial \indent\indent radius (\Rvir) at $z=0$.
\item[] C$'_3$: The satellite analog must have a subhalo mass $\geq 10^{9}$ \Msun  \indent\indent at $z=0$. 
\end{itemize}

A minimum subhalo mass of $10^9$ \Msun corresponds to $\sim$133 dark matter particles in Illustris-Dark.

\subsubsection{Prior 1}
\label{subsubsec:prior1}

Prior 1 is the sample of host halos that host exactly one massive satellite analog (criteria ${\bf C}$, Section \ref{subsubsec:massivesats}) and one or more low mass satellite analogs (criteria ${\bf C'}$, Section \ref{subsubsec:lowmasssats}). Approximately 92.8\% of host halos from the prior sample used in P17B satisfy these combined criteria. \emph{This subset of 18,236 host halos and the properties of their associated massive satellite analogs will be referred to as {\bf Prior 1} from here on.} Note that this prior is only trivially different from that selected in P17B. The LMC's properties will be used with Prior 1 to compute the mass of the MW.

\subsubsection{Prior 2}
\label{subsubsec:prior2}
 We find 87,598 subhalos that satisfy the low mass satellite analog criteria in a total of 18,236 unique host halos. Thus, many {\em host halos} harbor more than one low mass satellite analog, as expected. \emph{This set of host halos and the properties of their associated low mass satellite analogs will be referred to as {\bf Prior 2}.} Prior 2 will be used to compute the mass of the MW using the properties of all classical MW satellites less massive than the MCs.

Fig.~\ref{fig:priorscatterothersats} shows the distribution of latent properties ${\bm x}$ for a select set of subhalos in Prior 2. We have plotted the properties of subhalos residing in the bottom, middle, and top 5\% of the host halo mass distribution. Each panel indicates a pair of two parameters in ${\bm x} = [r, v_{\rm tot}, j]$ colored by the corresponding host halo mass. Overplotted are the observed properties ${\bm d} = [r^{\rm obs}, v^{\rm obs}_{\rm tot}, j^{\rm obs}]$ for all MW satellites listed in Table \ref{table:bayesparams}. The overlap between the observed properties and prior sample properties shows that the low mass satellite analogs accurately represent the true MW satellite properties. Without implementing any statistical techniques, the current properties of the classical MW satellites are most similar to those subhalos residing in host halos with masses $< 1.4\times 10^{12}$ \Msun.

The greatest number of low mass satellite analogs (by definition) hosted by any individual host halo in Prior 2 is 14 subhalos. Only 0.05\% of all halos considered in the prior sample host exactly 14 low mass satellite analogs. About 6\% of all host halos have 10 or more satisfactory subhalos within their virial radii. Typically, halos host between 2-5 low mass satellite analogs. See Appendix \ref{sec:appendix1} for more information on the abundance of subhalos within the virial radii of host halos in Prior 2.

The covariance between measurement errors of $r^{\rm obs}$ and $v_{\rm tot}^{\rm obs}$ is small, as shown in \citet{busha11}, so we treat these two measurements independently in the {\em instantaneous method}, which uses these two satellite properties to compute likelihoods for all halos in the prior sample. We consider the magnitude of specific orbital angular momenta separately in a parallel analysis, which we refer to as the {\em momentum method}. The most ideal candidates for our analysis are those residing at 100-300 kpc from the center of mass of their host halos since they are least affected by strong tides from the host (see Appendix \ref{sec:appendix2} for more details). 

\subsubsection*{M\MakeLowercase{ultiplicity of} H\MakeLowercase{ost} H\MakeLowercase{alos}}
In Prior 2, all host halos with more than one low mass satellite analog are counted towards the prior distribution of host halo masses multiple times. For example, if a host halo has a virial mass of \Mvir=$1.23 \times 10^{12}$ \Msun and hosts four low mass satellite analogs, its virial mass will appear four times in the list of halo masses that correspond to low mass satellite analogs. In P17B (and subsequently Prior 1), there is a one to one relation between host halos and massive satellite analogs since that prior was limited to hosts with just one massive satellite analog.

\subsection{Likelihood Functions for Low Mass Satellites}
\label{subsec:likelihoods}
For those host halos and low mass satellite analogs included in Prior 2, the physical parameters of interest are ${\bm \theta} =\{ {\bm x}, M_{\rm vir} \}$, where \Mvir is the virial mass of the corresponding host halo for any given subhalo. The parameters ${\bm x}$ are the latent, observable properties for satellite subhalos in Illustris-Dark. Fig. \ref{fig:priorscatterothersats} illustrates these for a fraction of the subhalos in the prior. The observational data $({\bm d})$ listed in Table~\ref{table:bayesparams} are measurements of the parameters ${\bm x}$, so if the measurement errors are zero, then ${\bm d=\bm x}$. Subsets of the physical parameters ${\bm x}$ are used in the different likelihood functions to estimate the mass of the MW.

The likelihood functions from P17B are altered such that they no longer rely on $v_{\rm max}^{\rm obs}$ due to the TBTF problem. We implement two methods to compute the likelihood of a given MW mass, as given below. \\  
\emph{Instantaneous} \begin{equation} \mathcal{L}({\bm x|\, \bm d})=  N(r^{obs} |\, r, \sigma_r^2) \times N(v_{tot}^{obs} |\, v_{tot}, \sigma_v^2), \label{eq:L4}  \end{equation} 
\emph{Momentum}\begin{equation}\mathcal{L}({\bm x|\, \bm d}) =  N(~j^{obs} |~j, \sigma_j^2) \label{eq:L5}  \end{equation}

All measured satellite properties are assumed to have Gaussian error. The posterior distribution for the virial halo mass of the MW using the properties of each low mass satellite is computed using Prior 2 and these likelihood functions via importance sampling. 

\subsubsection{Importance Sampling}
Bayes' theorem is written as
\begin{equation}\label{eq:bayes} 
 P({\bm x}, M_{\rm vir}|\,{ \bm d, C'}) \propto P({\bm d}| \, {\bm x}) \times P({\bm x}, M_{\rm vir}|\, {\bm C'}),
\end{equation}
where we denote the dependence on the prior selection criteria $\bm C'$. The left hand side is the posterior probability distribution.  $P({\bm x}, M_{\rm vir}|\, {\bm C')}$ represents the prior probability distribution and $P({\bm d}| \, {\bm x})$ is the likelihood (equivalent to $\mathcal{L}({\bm x|\, \bm d})$). 

A posterior probability density function (PDF) is then calculated by drawing a set of samples ($n$) from an importance sampling function. Here, we have chosen the importance sampling function to be the prior PDF (as in P17B). For each sample in $n$, an importance sampling weight proportional to the likelihood (using either Eq. \ref{eq:L4} or \ref{eq:L5}) is assigned. Using these weights, integrals that summarize the posterior PDF for halo mass are calculated as follows where the denominator is a normalization constant. 
 \begin{equation}
 \begin{split}
 \int f(\bm{\theta}) P({\bm x}, M_{\rm vir}|\, {\bm d, C'}) \, & d\bm{\theta} = \\
 & \frac{\int f(\bm{\theta})P({\bm d}|\,{\bm x})P({\bm x}, M_{\rm vir}| {\bm C'}) \, d\bm{\theta}}{\int P({\bm d}|\,{\bm x}) P({\bm x}, M_{\rm vir}|\, {\bm C'}) \, d\bm{\theta}} \\
&\approx \frac{\sum_j^n  f(\bm{\theta}_j)P({\bm d}|\,{\bm x}_j)}{\sum^n_j P({\bm d}|\,{\bm x}_j)}. 
\end{split}
\end{equation}

If $f(\theta)$ only depends on \Mvir, a representation of the marginal posterior PDF for \Mvir can be derived by computing Eq. \ref{eq:PDF} over a grid of potential halo mass values (i.e. using kernel density estimation). 
\begin{equation} \label{eq:PDF}
\begin{split}
 \int  f(M_{\rm vir}) P & (M_{\rm vir}|\, {\bm d, C'}) \, dM_{\rm vir} \\ 
= &\int \int  f(M_{\rm vir})\,  P({\bm x}, M_{\rm vir}|\, {\bm d, C'}) \, d{\bm x} \, dM_{\rm vir} \\
\approx & \frac{\sum_j^n  f(M_{\rm vir}^j) \, P({\bm d}| \,{\bm x}_j)}{\sum^n_j  P({\bm d}|\,{\bm x}_j)} \\
=&\sum_j^n  f(M_{\rm vir}^j) \, w_j \end{split}
\end{equation}
where $ w_i = P({\bm d}|\,{\bm x}_i)/\sum^n_j P({\bm d}|\,{\bm x}_j)$ are importance weights. Setting $f(\bm{\theta}) =$ \Mvir gives the posterior mean value for virial halo mass of the MW. For more details on the importance sampling technique, see Section 3.2 of P17B. 

Note that in practice all calculations are carried out in $\rm log_{10}$\Mvir (by directly replacing \Mvir with  $\rm log_{10}$\Mvir in all equations) because the posterior distribution of $\rm log_{10}$\Mvir is more roughly Gaussian than the posterior distribution of \Mvir, and therefore more easily summarized by a central value. Consequently, all results reported on a physical scale as \Mvir$=X^{+U}_{-L}$ \Msun should be interpreted on a log scale. For example, $\rm log_{10}$X is the posterior mean of $\rm log_{10}$\Mvir and [$\rm log_{10}$(X-L), $\rm log_{10}$(X+U)] is the 68\% credible interval in $\rm log_{10}$\Mvir. These summaries should not be naively translated to constraints on a linear scale as probability densities do not trivially transform under a nonlinear change of variables (see Jensen's inequality). The MW mass inferred by each satellite is discussed in Section \ref{subsec:indivresults}.

\subsection{A Statistical Approximation to Include Several Low Mass Satellites Simultaneously}
\label{subsec:statapprox}
Thus far, we have only outlined how to infer the virial mass of the MW using the properties of any individual satellite galaxy for which the proper motion has been measured. While these individual estimates are interesting on their own, such an analysis leads to the question: what if the phase space information of all satellites is used simultaneously to infer the mass of the MW? One might expect that additional information from multiple satellites will yield a more precise MW mass.

Below we outline a statistical approximation to simultaneously infer the MW's mass using the properties of all nine low mass satellites. An approximation is necessary to make a combined MW mass estimate since Prior 2 does not exclusively include only halos with approximately 10 low mass satellites each (see Section~\ref{subsec:prior}). To consider an arbitrary number of satellites with measurements $\{ \bm{d}_s\}$, Bayes' Theorem yields the joint posterior of host halo virial mass {$ \bm M_{\rm vir}$}:
\begin{equation}
\begin{split}
P( \{\bm{x}_s\},  M_{\rm vir} | \, & \{\bm{d}_s\}, \bm{C}') = \\
& \frac{P( \{\bm{d}_s\}  | \, \{\bm{x}_s\}) \times P( \{ \bm{x}_s \}, M_{\rm vir} | \, \bm{C}')}{P(\{\bm{d}_s\} | \, \bm{C}')}
\end{split}
\end{equation}
where $s = 1 \ldots N_{\rm sat}$ and  $N_{\rm sat}$ is the total number of satellites considered. Because the measurements of each satellite are independent from the others, the likelihood can be written
\begin{equation}
P( \{\bm{d}_s\} | \, \{\bm{x}_s\}) = \prod_{s=1}^{N_{\rm sat}} P( \bm{d}_s | \, \bm{x}_s ).
\end{equation}
Next, we note that the prior factor can be written, using the definition of conditional probability,
\begin{equation}
P( \{ \bm{x}_s \}, M_{\rm vir} | \, \bm{C}') =   P(M_{\rm vir} | \, \bm{C}') \times P( \{ \bm{x}_s \} | \, M_{\rm vir},  \bm{C}').
\end{equation}
If we make the \emph{naive Bayes} assumption that, given the mass $M_{\rm vir}$, the satellites properties $\{\bm{x}_s\}$ are conditionally independent, we have
\begin{equation}
P( \{ \bm{x}_s \} | \, M_{\rm vir},  \bm{C}') =  \prod_{s=1}^{N_{\rm sat}} P( \bm{x}_s  | \, M_{\rm vir},  \bm{C}'),
\end{equation}
and therefore,
\begin{equation}
P( \{ \bm{x}_s \}, M_{\rm vir} | \, \bm{C}') = P(M_{\rm vir} | \, \bm{C}') \times \prod_{s=1}^{N_{\rm sat}} P( \bm{x}_s  | \, M_{\rm vir},  \bm{C}').
\end{equation}
Putting this all together, we have
\begin{equation}
\begin{split}
P(& \{\bm{x}_s\},  M_{\rm vir} |  \, \{\bm{d}_s\}, \bm{C}') =  \\
& \frac{ \Big[ \prod_{s=1}^{N_{\rm sat}} P( \bm{d}_s | \, \bm{x}_s)  P(  \bm{x}_s  | \, M_{\rm vir}, \bm{C}') \Big] \times P(M_{\rm vir} | \, \bm{C}')}{P(\{\bm{d}_s\} | \, \bm{C}')}.
\end{split}
\end{equation}
The individual prior factors can be written
\begin{equation}
P(  \bm{x}_s  | \, M_{\rm vir}, \bm{C}') = \frac{P(  \bm{x}_s, M_{\rm vir} | \bm{C}') }{P(M_{\rm vir} | \bm{C}')},
\end{equation}
so the joint posterior becomes:
\begin{equation}
\begin{split}
& P(\{\bm{x}_s\}, M_{\rm vir} | \, \{\bm{d}_s\},  \bm{C}') =  \\
& \frac{ \Big[ \prod_{s=1}^{N_{\rm sat}} P( \bm{d}_s | \, \bm{x}_s)  P(  \bm{x}_s,  M_{\rm vir} |\,\bm{C}') \Big] \times P(M_{\rm vir} | \, \bm{C}')^{1-N_{\rm sat}}} {P(\{\bm{d}_s\} | \, \bm{C}')}.
\end{split}
\end{equation}
Next we notice that the posterior given the data for a single satellite $s$, is
\begin{equation}
P( \bm{x}_s, M_{\rm vir} | \, \bm{d}_s, \bm{C}') =  \frac{P( \bm{d}_s | \, \bm{x}_s)  P(  \bm{x}_s,  M_{\rm vir} |\,\bm{C}') }{P(\bm{d}_s | \, \bm{C}')},
\end{equation}
so the joint posterior is
\begin{equation}
\begin{split}
P( \{\bm{x}_s\}, M_{\rm vir} | \, \{\bm{d}_s\}, \bm{C}') &= \Big[ \prod_{s=1}^{N_{\rm sat}} P( \bm{x}_s, M_{\rm vir} | \, \bm{d}_s, \bm{C}') \Big] \times  \\ P(M_{\rm vir} | \, \bm{C}')^{1-N_{\rm sat}}
&\times \frac{\prod_{s=1}^{N_{\rm sat}}  P( \bm{d}_s | \, \bm{C}')}{P( \{\bm{d}_s\} | \, \bm{C}')}.
\end{split}
\end{equation}
Integrating out $\{ \bm{x}_s \}$, we find the marginal posterior for the mass given all of the satellites' data.
\begin{equation}\label{eqn:marg_post_all_at_once}
\begin{split}
P(  M_{\rm vir} | \, \{\bm{d}_s\}, \bm{C}') &= \Big[ \prod_{s=1}^{N_{\rm sat}} P( M_{\rm vir} | \, \bm{d}_s, \bm{C}') \Big] \times \\ 
P(M_{\rm vir} | \, \bm{C}')^{1-N_{\rm sat}}
&\times \frac{\prod_{s=1}^{N_{\rm sat}}  P( \bm{d}_s | \, \bm{C}')}{P(\{ \bm{d}_s \}| \, \bm{C}')}
\end{split}
\end{equation}
The last factor is a normalization constant that does not depend on the parameters.  Since we already know how to calculate $P( M_{\rm vir} | \, \bm{d}_s, \bm{C}')$ for one satellite at a time using Eq. \ref{eq:PDF}, a useful expression for the marginal posterior is:
\begin{equation} \begin{split}
P(  M_{\rm vir} | \, \{\bm{d}_s\}, \bm{C}') \propto \Big[ & \prod_{s=1}^{N_{\rm sat}} P( M_{\rm vir} | \, \bm{d}_s, \bm{C}') \Big]  \times \\ & P(M_{\rm vir} | \, \bm{C}')^{1-N_{\rm sat}}.
\label{eq:final}
\end{split}
\end{equation}

By writing the marginal posterior distribution in this form, the multiplicity from including Prior 2 $ N_{\rm sat}$ times is eliminated. Again, all calculations are computed in $\rm log_{10}$\Mvir as noted in Section \ref{subsec:likelihoods}. Results for the combined MW mass estimates using this statistical approximation are discussed in Section \ref{subsec:comboresults}. 

\subsection{The Conditional Independence Assumption and Computing the Joint Posterior Distribution}
\label{sec:condindep}

\begin{figure*}
\begin{center}
\includegraphics[scale=0.5, trim = 5mm 0mm 12mm 0mm]{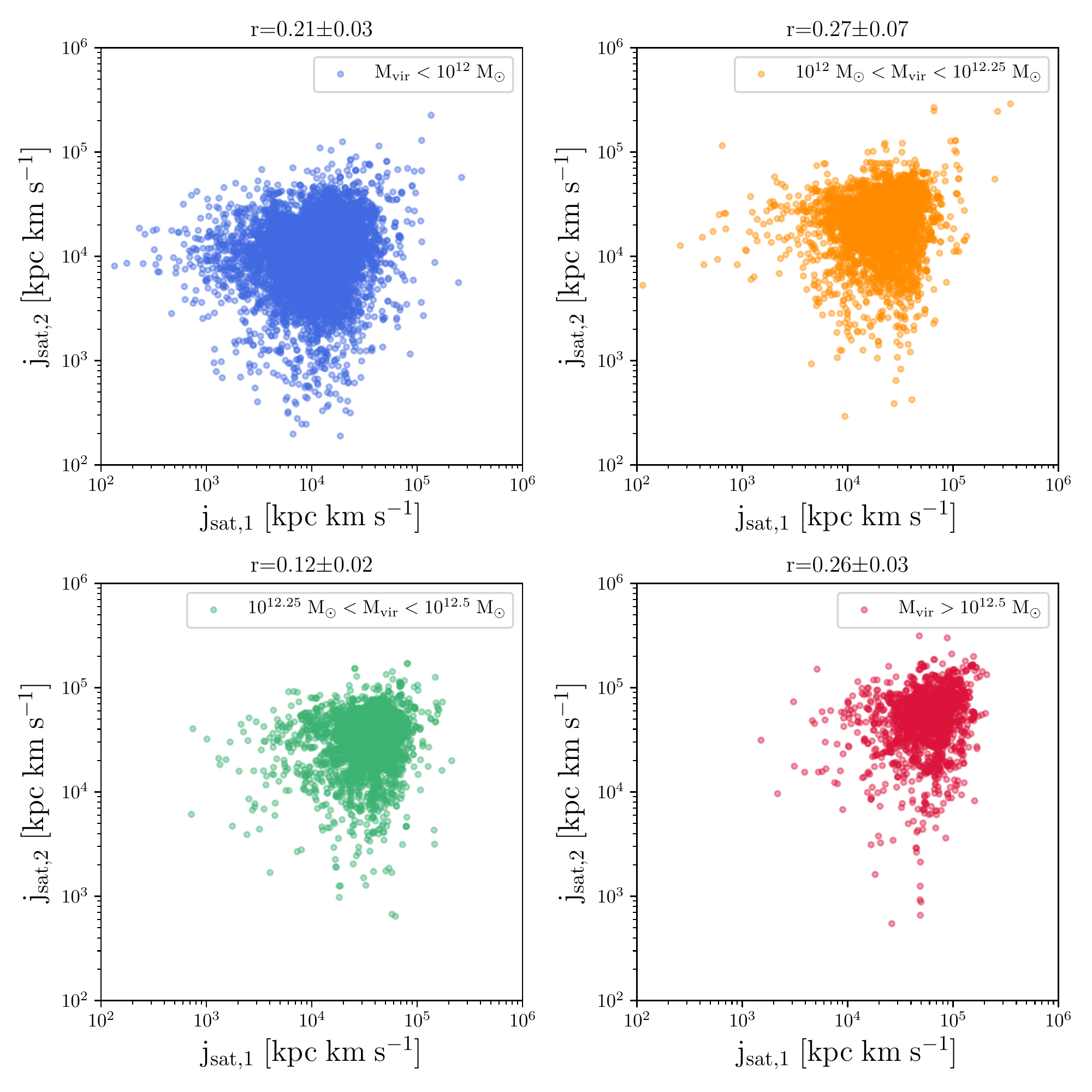}
\caption{For all hosts in Prior 2 that host at least two low mass satellite analogs, the magnitude of the specific orbital angular momentum of the first (more massive) satellite is plotted against that of the second satellite. The sample is split into four bins based on host halo mass. Pearson's correlation coefficient (PCC) for each subsample is denoted by the $r$ value above each panel. The PCC for the whole sample population is $r_{all}=0.56\pm 0.01$. Errors are computed using bootstrap re-sampling. There is only a weak correlation between $j_{\rm sat, 1}$ and $j_{\rm sat,2}$, indicating that the conditional independence assumption is reasonable.
\label{fig:conditionalindep}}
\end{center}
\end{figure*}

The conditional independence assumption states that the properties of a subhalo in a given host halo are independent from those of another subhalo in the same host halo for each given value of host halo mass. Here, we demonstrate the validity of this assumption for our statistical framework. 

The conditional independence assumption requires that the correlation based on $P( j_{\rm sat,1}, j_{\rm sat,2} | M_{\rm vir} )$, for example, is zero for all \Mvir where $j_{\rm sat, 1}$ and $j_{\rm sat,2}$ are the values of total specific orbital angular momentum for the first two low mass satellite analogs in each host halo (where at least two analogs exist). In Fig. \ref{fig:conditionalindep}, we plot $j_{\rm sat, 1}$ versus $j_{\rm sat,2}$ for the fraction of Prior 2 as four subsamples split by host halo mass. The assumption should hold for all values of host halo mass. The Pearson correlation coefficient ($r$) between $j_{\rm sat,1}$ and  $j_{\rm sat,2}$ for each subsample yields values between 0.12-0.27 with uncertainties < 0.07. 

Physically, this weak correlation shows that there is reasonable scatter amongst satellite specific orbital angular momenta for the same host halo. This suggests that satellites' angular momenta vectors are not set by the strength of the large scale tidal field but rather by more complex processes such as varying accretion and orbital histories even for satellites orbiting the same host. Such values suggest there is only a weak 2-point correlation and therefore that the conditional independence is a useful and reasonable approximation. The conditional independence assumption actually requires that the N dimensional joint distribution factors as:
\begin{equation} 
\begin{split} P(& j_{\rm sat,1} , j_{\rm sat,2} ... ,  j_{\rm sat,N} | M_{\rm vir} ) = \\ 
 P( & j_{\rm sat,1} | M_{\rm vir} ) \times P(j_{\rm sat,2} | M_{\rm vir} ) \times ... \times P(j_{\rm sat,N} | M_{\rm vir}. )
\end{split}\end {equation}
However, we have only demonstrated this for N=2 in Fig. \ref{fig:conditionalindep}, as it is difficult to rigorously show this for N > 2.

Calculating Eq. \ref{eq:final} directly can sometimes lead to numerical underflow or overflow errors. To prevent this, we compute the logarithm of Eq. \ref{eq:final}, apply it to our data, and exponentiate to retrieve the final results. This strategy is successful in a majority of cases unless the product of all satellite kernel density estimate (KDE) posteriors and the KDE estimate of the prior in Eq. \ref{eq:final} both approach zero in a numerically unstable way. In such cases, the limiting edge effects must be carefully considered to produce a stable result. Such numerical caveats will be unnecessary when more advanced high resolution simulations are available.

\section{MW Mass Results Using the Classical Dwarf Satellites}
\label{sec:results}

We now infer the mass of the MW using both the instantaneous and momentum likelihood methods with the properties of each individual dwarf satellite and the Illustris-Dark cosmological simulation. While we already demonstrated in P17B that the momentum method is more reliable as a function of time and satellite orbital phase, we will report results from both likelihood functions for comparison. We also include the results for the ensemble MW mass estimates using all classical satellites.

In what follows, we provide mass estimates of the MW using the LMC's orbital properties and Prior 1. However, this mass estimate cannot be combined with those resulting from the lower mass satellite galaxies as the two sets of results are computed from two different prior samples (Prior 1 vs. Prior 2). 

\subsection{MW Mass Estimates from Individual Low Mass Dwarf Satellites}
\label{subsec:indivresults}
\begin{figure*}
\begin{center}
\includegraphics[scale=0.6]{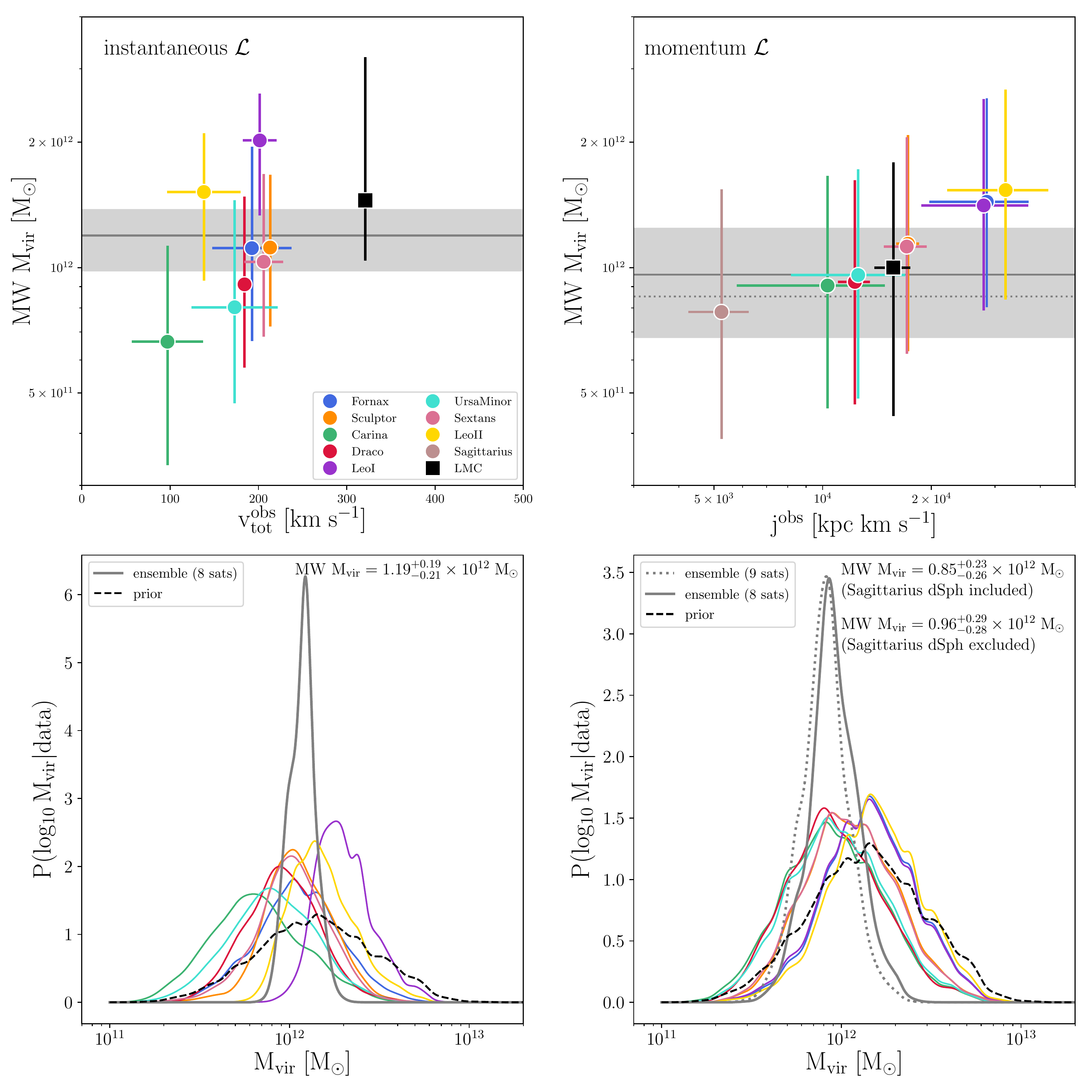}
\caption{{\em Top panels:} Summaries of the posterior distributions for MW halo mass using Prior 2 and the observed properties of the classical low mass MW satellites with proper motions. Markers with error bars indicate the posterior mean halo mass and the halo mass included in the 68\% credible interval. The left panel shows the results from the instantaneous likelihood method versus the total velocity of each satellite today relative to the MW. The right panel shows the results from the momentum likelihood method versus the magnitude of the current specific orbital angular momenta. The gray shaded regions indicate the 68\% credible interval for MW halo mass using the ensemble of low mass satellites (i.e. excluding the MCs and Sgr dSph).  {\em Bottom panels:} The posterior distributions in MW halo mass for each individual low mass satellite are shown as colored curves. The black dashed lines indicate the probably distribution of the prior sample given equal likelihood weights. The solid gray curves represent the ensemble posterior distributions (excluding Sgr dSph) and the dotted gray curve in the right panel is the ensemble posterior including Sgr dSph. The resulting posterior mean masses are listed in the top right of each panel. All calculations are computed in $\rm log_{10}(M_{\rm vir})$. See Section \ref{subsec:comboresults} for details on the alignment of the ensemble posterior distributions with the individual posterior distributions.
\label{fig:summaryothersats}}
\end{center}
\end{figure*}

\begin{table}
\caption{Summary Statistics for MW Mass Estimates}
\label{table:results}
\begin{center}
\begin{tabular}{ccc} \hline \hline
 & Instantaneous & Momentum\\ 
 & \Mvir [$\rm 10^{12}\; M_{\odot}$] & \Mvir [$\rm 10^ {12}\; M_{\odot}$] \\ \hline \hline
 
LMC & $1.45^{+1.75}_{-0.41}$ & $1.0^{+0.79}_{-0.56}$\\ \hline 
Leo II & $1.52^{+0.58}_{-0.59}$ & $1.54^{+1.14}_{-0.7}$ \\ 
Fornax &  $1.11^{+0.84}_{-0.45}$ & $1.44^{+1.11}_{-0.63}$\\ 
Leo I & $2.02^{+0.60}_{-0.69}$ & $1.41^{+1.13}_{-0.62}$ \\ 
Sculptor & $1.12^{+0.55}_{-0.39}$ & $1.14^{+0.94}_{-0.51}$ \\ 
Sextans & $1.03^{+0.64}_{-0.35}$ & $1.12^{+0.93}_{-0.50}$ \\ 
Ursa Minor & $0.80^{+0.65}_{-0.33}$& $0.96^{+0.76}_{-0.48}$ \\ 
Draco & $0.91^{+0.57}_{-0.34}$ & $0.92^{+0.70}_{-0.45}$ \\ 
Carina & $0.66^{+0.46}_{-0.33}$ & $0.91^{+0.76}_{-0.45}$\\ \hline \hline
$\rm {\bf Ensemble\; (8\;sats)}$ & $1.19^{+0.19}_{-0.21}$ & $ 0.96^{+0.29}_{-0.28}$\\  \hline 
Sagittarius dSph & -- & $0.78^{+0.76}_{-0.39}$ \\ \hline \hline
$\rm {\bf Ensemble\; (9\; sats)}$ & -- &  $0.85^{+0.23}_{-0.26}$\\  \hline
\end{tabular}
\end{center}
\tablecomments{The posterior mean and 68\% credible interval in halo mass for the MW using the properties of each of the nine low mass satellites and the LMC. The MW halo mass computed using host halos in Prior 2 and the properties of eight (nine) low mass satellites simultaneously is given in the third to last row. The MW mass derived using the LMC was calculated with Prior 1 (first row). The LMC is excluded from the ensemble mass estimates since these calculations use fundamentally different prior samples (Prior 1 vs. Prior 2). The SMC's properties are not considered due to the low frequency of LMC-SMC analog pairs in Illustris-Dark.}

\end{table}

\subsubsection{Instantaneous Likelihood}

The top left panel of Fig.~\ref{fig:summaryothersats} shows the MW mass estimates and associated uncertainties for each low mass satellite using the instantaneous likelihood method. The inferred MW masses are plotted against the total velocity of each satellite relative to the MW today (Table~\ref{table:bayesparams}, Column 4). Error bars indicate the MW's posterior mean mass included in the 68\% credible interval of the posterior distribution for host halo mass. The colored lines in the bottom left panel of Fig.~\ref{fig:summaryothersats} show the corresponding posterior distributions. All results are also listed in Table~\ref{table:results}. The LMC is indicated by a black square in Fig.~\ref{fig:summaryothersats}. 

The results of the instantaneous method display a fairly large scatter (standard deviation of $\sim0.14$ dex) among the posterior mean mass estimates for the MW using each individual satellite. The lowest value for the MW's mass comes from Carina, which suggests a posterior mean MW mass of only $0.66\times10^{12}$ \Msun. As expected, the highest MW mass estimate ($2.02\times10^{12}$ \Msun) with the instantaneous method comes from Leo I. This mass is consistent with the bound mass arguments given by BK13 and the mass inferred by the radial timing-argument, but lower than that given by the non-radial timing argument \citep{liwhite08, sohn13}. Carina is on a low energy, fairly circular orbit about the MW \citep{pasetto11}, whereas Leo I is on a high energy orbit and has recently completed its first pericentric approach \citep{ sohn13}. The difference in orbital energies suggests a strong correlation between energy and halo mass. 

As any one individual satellite can produce a MW mass estimate that could be misleading given its orbital energy, this warrants combining as many satellites as possible to infer the mass of the MW in tandem. The instantaneous method essentially reproduces expectations for the MW's mass from traditional methods such that Leo I pushes the MW's mass to higher values and satellites on more circular orbits that experience little to no dynamical friction prefer lower MW halo masses.  

The Sagittarius dSph has been omitted from the results for the instantaneous method as the effective sample size reduces to just a few during the importance sampling step given its unique combination of position and velocity. Upon examining the cumulative number density of subhalos in Prior 2 at any given distance relative to their respective host, we find that the minimum host-satellite separation is $\sim$ 30 kpc. Therefore, very few subhalos are found at the separation of Sagittarius (20 kpc) in Illustris-Dark. This problem may be two-fold as the gravitational smoothing length of Illustris-Dark may not be able to sufficiently resolve distinct halos at these small separations and there may also be a depletion of subhalos in the inner regions of halos due to tidal disruption, especially for subhalos on radial orbits \citep{garrisonkimmel17}.

While only $\sim$20\% of the known MW satellites \citep[e.g.][]{dwagner15} are considered in this analysis, the individual estimates using the instantaneous method already demonstrate a factor of three scatter in MW halo mass, even larger than that in the literature to date. The instantaneous method is not recommended if unbiased MW mass estimates are desired.

\subsubsection{Momentum Likelihood}
\label{subsubsec:momentumresults}
MW mass estimates for the momentum method are plotted in the top right panel of Fig.~\ref{fig:summaryothersats} against the current magnitude of specific orbital angular momentum for each satellite. The bottom right panel shows the corresponding posterior distributions. In general, we see a similar trend amongst the low mass satellites as we did for the LMC and M33 analysis in P17B -- satellites that have a higher total specific orbital angular momentum value (see Table~\ref{table:bayesparams}) infer higher MW masses. The same trend is not strictly true for $\rm v_{tot}^{obs}$ (Fig. \ref{fig:summaryothersats}, top left panel) or $\rm r^{obs}$. Overall, the scatter among individual MW mass estimates using the momentum method is far better constrained to a range of $0.78\times10^{12}$ \Msun (Sagittarius dSph) -- $1.54\times10^{12}$\Msun (Leo II), in agreement with the current factor of two spread in MW mass. Excluding Sagittarius dSph, the scatter of our results is narrowed even further to $0.91-1.54\times 10^{12}$ \Msun. Comparing the standard deviation across individual estimates from the instantaneous method and the momentum method gives $\sim0.14$ dex versus $\sim0.09$ dex, clearly demonstrating the improvement that the momentum method provides if satellites are used individually.

In P17B, we demonstrated that using instantaneous properties like position and velocity skew the resulting MW mass estimates because they change significantly with time. The specific orbital angular momentum vector of Sagittarius dSph has likely undergone significant changes since its infall into the halo of the MW due to the tidal stripping of its stars and subsequent formation of the Sagittarius stellar stream \citep[see][]{belokurov14}. It is therefore unsurprising that \citet{gibbons14} and our analysis yield such low MW masses using the Sagittarius dSph (or its associated stream). The MW mass inferred by the Sagittarius dSph should therefore be taken with caution. 

Contrary to previous studies, Leo II, rather than Leo I, puts limits on the MW's mass. Leo II has the highest specific orbital angular momentum of the satellites considered in this analysis, so it brackets the upper end of the MW's plausible mass range.

Satellites on the most extreme orbits, where they have only made one pericentric passage about the MW (Leo I or the LMC), also show the biggest deviations between the posterior mean masses estimates using the instantaneous versus the momentum method. The fact that the MW mass estimates vary so drastically for these satellites provides further evidence that the combination of position and velocity is not reliable for recovering the mass of the MW accurately, and cautions against using one satellite alone to infer the MW's mass. 

We conclude that the momentum method is not only a more consistent estimator of host halo mass as a function of time and orbital evolution but it is also a more consistent method for determining the mass of a host halo given the 6D phase space information for a \emph{population} of satellite galaxies. The momentum method directly correlates inferred host halo mass and satellite specific orbital angular momentum, thereby distinguishing between halos that can host a population of satellites exhibiting a given observationally constrained distribution of specific orbital angular momenta and those that cannot. The correlation between the distribution of specific orbital angular momentum for a population of satellites and host halo mass will be discussed further in Section~\ref{sec:discussion}. 

While the momentum method for low mass satellites only considers one physical property associated with each satellite -- the magnitude of the specific orbital angular momentum -- it encompasses some 6D phase space information of each satellite too as $j = |{\bm r} \times {\bm v}|$, where ${\bm r}$ and ${\bm v}$ are 3D vectors. The direction of the specific orbital angular momentum method is not utilized in this method, as given. In principle, such information could be incorporated into the momentum likelihood function, but this is beyond the scope of this work. 

\subsubsection{Caveats for the Individual MW Mass Estimates}
In P17B, we carried out an additional bootstrap analysis to address the sampling noise associated with our technique. This is especially important for satellites in unique orbital configurations, like the LMC and the Sagittarius dSph. Since Prior 2 is significantly larger than that considered in P17B, the effective sample size does not rapidly decrease to zero for most of the classical MW satellites (except Sagittarius dSph) considered. We also showed in Section 5.2 of P17B that smaller measurement errors can improve host halo mass estimates, but when the precision is already as low as a few percent (as is the case for the LMC, Sculptor, Draco), the change in results is insignificant. The Gaia mission will be able to reduce the measurement errors on proper motions and derived quantities for the nearest classical satellite galaxies \citep{vdMsahlmann}, further narrowing MW mass estimates using those satellites.

The barrier to achieving very high precision MW mass estimates for the individual satellites is the irreducible uncertainty owing to cosmic variance (Section 5.3, P17B), or the intrinsic correlation between host halo mass and satellite dynamics. We expect that the most significant improvements to the precision of MW mass estimates will therefore arise from using the properties of several satellites simultaneously. 

\subsection{MW Mass Estimates from the Ensemble of Low Mass Dwarf Satellites}
\label{subsec:comboresults}

Using the statistical approximation outlined in Section \ref{subsec:statapprox}, the most probable MW mass resulting from the ensemble of low mass classical satellites and each of the likelihood methods are represented by the gray shaded regions in the top panels of Fig.~\ref{fig:summaryothersats} and by the gray lines in the bottom panels of Fig.~\ref{fig:summaryothersats}. 

For the instantaneous method, the combination of eight low mass satellites (Sagittarius dSph excluded) yields an ensemble MW mass estimate of $\rm M_{vir, MW}= 1.19^{+0.19}_{-0.21}\times 10^{12}$ \Msun ($\rm log_{10}(M_{vir}/M_{\odot})=12.08^{+0.06}_{-0.09}$). The ensemble MW halo mass resulting from the momentum likelihood using the same eight satellites is $\rm M_{vir, MW}= 0.96^{+0.29}_{-0.28}\times10^{12}$ ($ \rm log_{10}(M_{vir}/M_{\odot})=11.98^{+0.11}_{-0.15}$). When Sagittarius dSph is included in the combined mass estimate using the momentum method, the mass of the MW decreases to $\rm M_{vir, MW}= 0.85^{+0.22}_{-0.26}\times 10^{12}$ \Msun ($ \rm log_{10}(M_{vir}/M_{\odot})=11.93^{+0.11}_{-0.16}$) with a larger uncertainty. The larger measurement errors for the observed $j$ values results in a wider 68\% credible interval for the momentum method compared to the instantaneous method, but all results are narrower than the current factor of two spread in mass and far more precise than those predicted by any one satellite. 

Note that the posterior distributions for the ensemble mass estimates in Fig. \ref{fig:summaryothersats} are slightly shifted relative to the posterior distributions for each individual satellite. This is because the ensemble posterior mass distributions are calculated by dividing out the multiplicity of the prior, whereas the individual posteriors still include one instance of the prior. If instead all of the individual posterior distributions were multiplied together to form the joint posteriors, they would align with the individual posterior distributions.

Recall that the SMC was excluded from the prior selection criteria (see Section \ref{subsec:prior}) and therefore its specific orbital angular momentum ($j_{\rm SMC} = 13,209 \pm 2,067$ kpc \kms) is not included in the ensemble mass estimates. While the SMC is not explicitly accounted for in this analysis, we expect that it is unlikely to significantly change the results since its specific orbital angular momentum lies approximately between that of Ursa Minor and the LMC.

Our ensemble mass estimates suggest that Leo I could be bound to the MW within the associated credible intervals. According to the upper 68\% credible interval when the Sagittarius dSph is excluded from our momentum method results, MW masses between $0.96-1.25 \times 10^{12}$ \Msun (\Rvir > 260 kpc) suggest that Leo I is bound to the MW. However, these results are preliminary since we only use the phase space information for 20\% of all known MW satellite galaxies. These conclusions should be revisited as the observational data set increases.

We conclude that sampling the full range of specific orbital angular momentum for the observed MW satellite population provides the most reliable mass estimate as it includes much of the available 6D phase space information in an unbiased fashion. Our results are in good agreement with the complementary work of \citet{li17}. They have used nine cosmological zoom simulations and a scaling method with the angular momentum and energy distribution of the classical MW satellites to conclude a MW mass of $1.3 \times 10^{12}$ \Msun with a $\sim$40\% error. They are also consistent with independent mass estimation methods such as those presented in \citet{kafle12,kafle14}, which use observations of the stellar halo and blue horizontal branch stars to estimate the MW's mass. 

\subsubsection{Tests and Caveats for Ensemble MW Mass Estimates}
To test whether the statistical approximation outlined in Section \ref{subsec:statapprox} yields accurate results when the individual posteriors of the classical MW satellites are combined via a statistical approximation, we estimated the mass of 100 random halos from Illustris-Dark. By comparing the estimated masses to the true host halo mass for these 100 halos, we can assess whether our method accurately recovers host halo mass. For this calculation, all subhalos in the 100 randomly chosen host halos were assigned a 20\% measurement error on the magnitude of their specific orbital angular momenta. For the true MW satellites, measurement errors range from about 7\% up to about 40\%. We find that in approximately 90\% of these halos, the true mass is contained within two posterior standard deviations of the posterior mean (in dex). Ideally, this would be true for 95\% of the halos but we expect that this 5\% deficit will disappear when more suitable simulations are available. Similar results are recovered when a 10\% measurement error is applied to the properties of all subhalos in these 100 test halos.

Note that the MW mass estimates resulting from the ensemble of classical satellites in this work are preliminary. Currently, Fig.~\ref{fig:summaryothersats} only shows the results for about 20\% of all known MW satellites. Recent work suggests that this satellite population is less than 10\% of the total number of satellites predicted around the MW \citep[see][]{tollerud11, newton17}. When the proper motions of additional low mass satellites, such as ultra-faint dwarfs, and a large volume simulation that can resolve analogs of ultra-faints become available, we can rigorously test the limits of this method. However, we expect that the combined MW mass estimate calculating using N satellites will eventually plateau due to cosmic variance. See \citet{li17} for predictions in the context of satellite galaxies and \citet{wang17} on how the number of independent phase-space structures contributes to mass uncertainties.

\section{Discussion}
\label{sec:discussion}

\begin{figure}
\begin{center}
\includegraphics[scale=0.58, trim = 10mm 0mm 0mm 0mm]{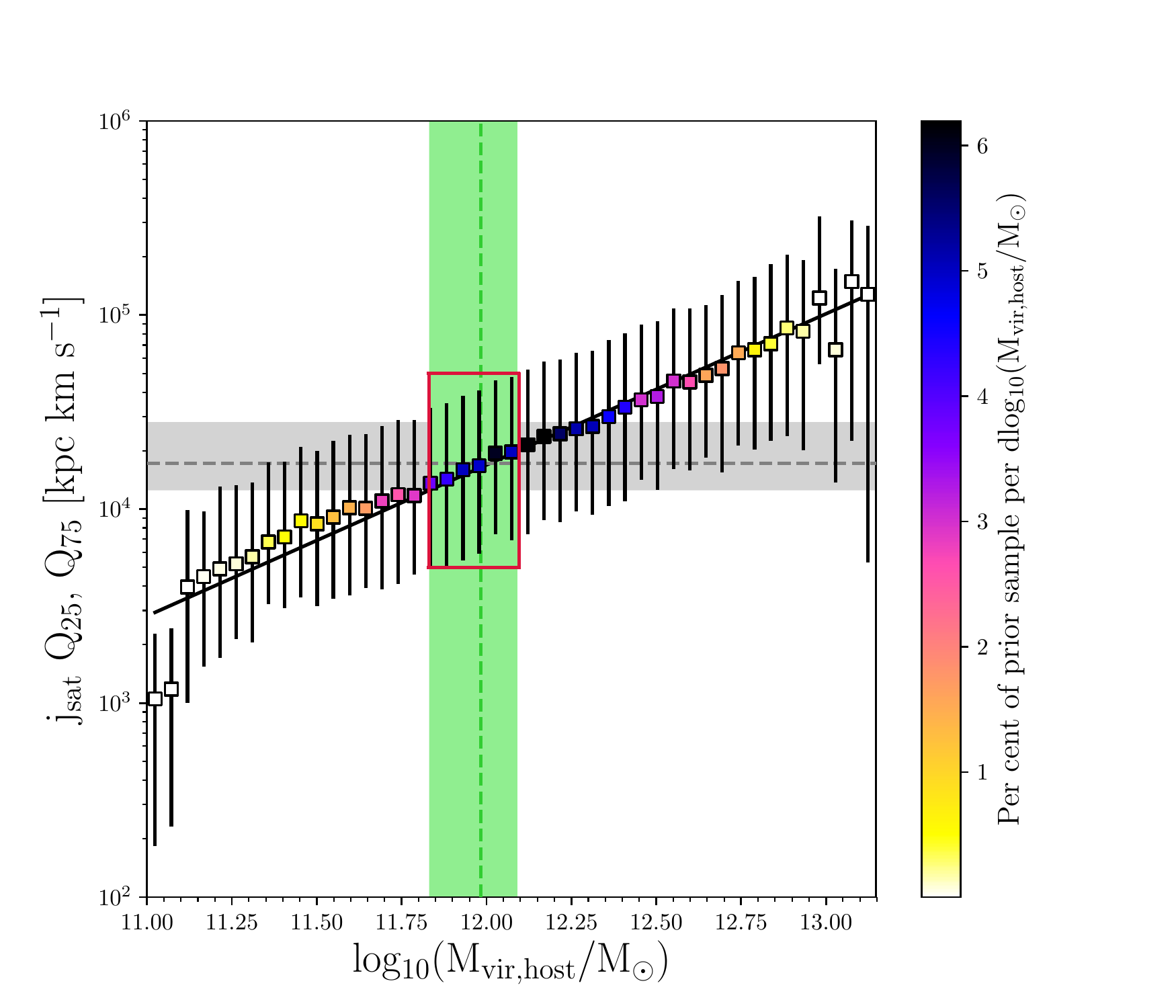}
\caption{The median value of specific orbital angular momentum for the low mass satellite analogs in Prior 2 binned by host halo virial mass. The error bars show the extents of the 25th and 75th percentiles for specific orbital angular momentum in each mass bin. The gray dashed line and shaded region show the median specific angular momentum value of eight low mass MW satellites (Sagittarius dSph excluded) and the extents of corresponding quartiles. The green dashed lines and shaded region represent the posterior mean mass of the ensemble estimate using the momentum likelihood and the corresponding 68\% credible interval ($\rm log_{10}(M_{vir}/M_{\odot})=11.98^{+0.11}_{-0.15}$). The color map indicates the percent of low mass subhalos in Prior 2 that fall in each host mass bin. The highest percentages of subhalos reside in host halos with masses log$\rm_{10}$(\Mvir/\Msun) $\approx$ 11.8-12.3, consistent with the properties of the classical dSphs and the results reported in this work. The black solid line is the line of best fit. When the specific orbital angular momentum values of the LMC analogs are added to the data sample shown here, the overall relation between the specific orbital angular momentum distribution and halo mass still holds. The red box indicates the expected $j$ values for ultra-faint dwarf galaxies.
\label{fig:mvirjrelation}}
\end{center}
\end{figure}

In Section~\ref{subsubsec:momentumresults}, we note the strong correlation between the specific orbital angular momenta distribution of the observed satellite population and the MW mass inferred by those satellites. Naturally, this leads to the question: is there a correlation between host halo mass and the distribution of specific orbital angular momentum for a population of subhalos? We use simulated analogs from Illustris-Dark to explore if this intrinsic relationship exists and how it might inform our knowledge of the MW's true mass. 

\subsection{The Halo Mass-Specific Angular Momenta Distribution Relation in the Presence of a Massive Satellite}

Fig.~\ref{fig:mvirjrelation} shows the distribution of total specific orbital angular momenta for subhalos in Prior 2 (Section \ref{subsubsec:prior2}) binned by their corresponding host halo mass. The specific orbital angular momenta shown are the median values for each host halo mass bin and the error bars indicate the extents of the 25th and 75th quartiles. The gray shaded area shows the median and quartiles of the specific orbital angular momentum distribution for the population of classical MW satellites considered in this work. 

The high concentration of subhalos residing in halos with masses between $\rm log_{10}(M_{vir}/M_{\odot})\approx 11.8-12.3$ suggests that this is the most typical MW halo mass independent of our results from Section \ref{sec:results}. This mass range is in agreement with the frequentist MW mass predictions from \citet{patel17a} using the energetics of LMC analogs in Illustris-Dark. Note that this region is also coincident with the gray shaded area, which represents the true distribution of orbital angular momentum for eight of the classical MW satellites. The green dashed line and shaded region represent the posterior mean and 68\% credible interval included in the ensemble MW mass estimate from the momentum method (see Table \ref{table:results}, last row). The black solid line is the line of best fit with a slope of $m=0.779\pm0.0280$ and an intercept of $b=-5.116\pm0.339$ in units of dex (i.e. the fits are calculated using $\rm log_{10}(j_{sat})$ and $\rm log_{10}(M_{vir}/M_{\odot})$).

When we further explore if the specific orbital angular momenta-host halo mass trend is correlated with the orbital history of the massive satellite analogs, we find that the linear relationship shown in Fig.~\ref{fig:mvirjrelation} is generally unaffected. The subhalos residing in halos whose massive satellite analogs have crossing times{\footnote{Crossing time is defined as the first time a subhalo crosses the time-evolving virial radius of its host halo, moving inwards. See \citet{patel17a} for more details.} less than 4 Gyr ago (first infall scenarios) and those hosting massive satellite analogs with crossing times more than 4 Gyr ago (multiple passages about their hosts) exhibit slopes in agreement with the best fit line in Fig.~\ref{fig:mvirjrelation} (black solid line).

Using Fig. \ref{fig:mvirjrelation}, we can also make predictions for the specific orbital angular momenta of ultra-faint dwarf galaxies. The red box, which encompasses the region of highest subhalo abundance (darkest blue squares), our MW mass estimate (green shaded region), and the observed distribution in specific orbital angular momentum of satellites (gray shaded region) warrants the prediction that ultra-faint dwarfs should generally exhibit $5 \times10^3$ kpc \kms $<  j^{obs}  < 5 \times 10^4$ kpc \kms if the MW's true mass is $\sim10^{12}$ \Msun. Upcoming observations of ultra-faint dwarf galaxies will be some of the most interesting in this regard given their expected abundance.

\begin{figure}
\begin{center}
\includegraphics[scale=0.58, trim = 10mm 0mm 0mm 0 mm]{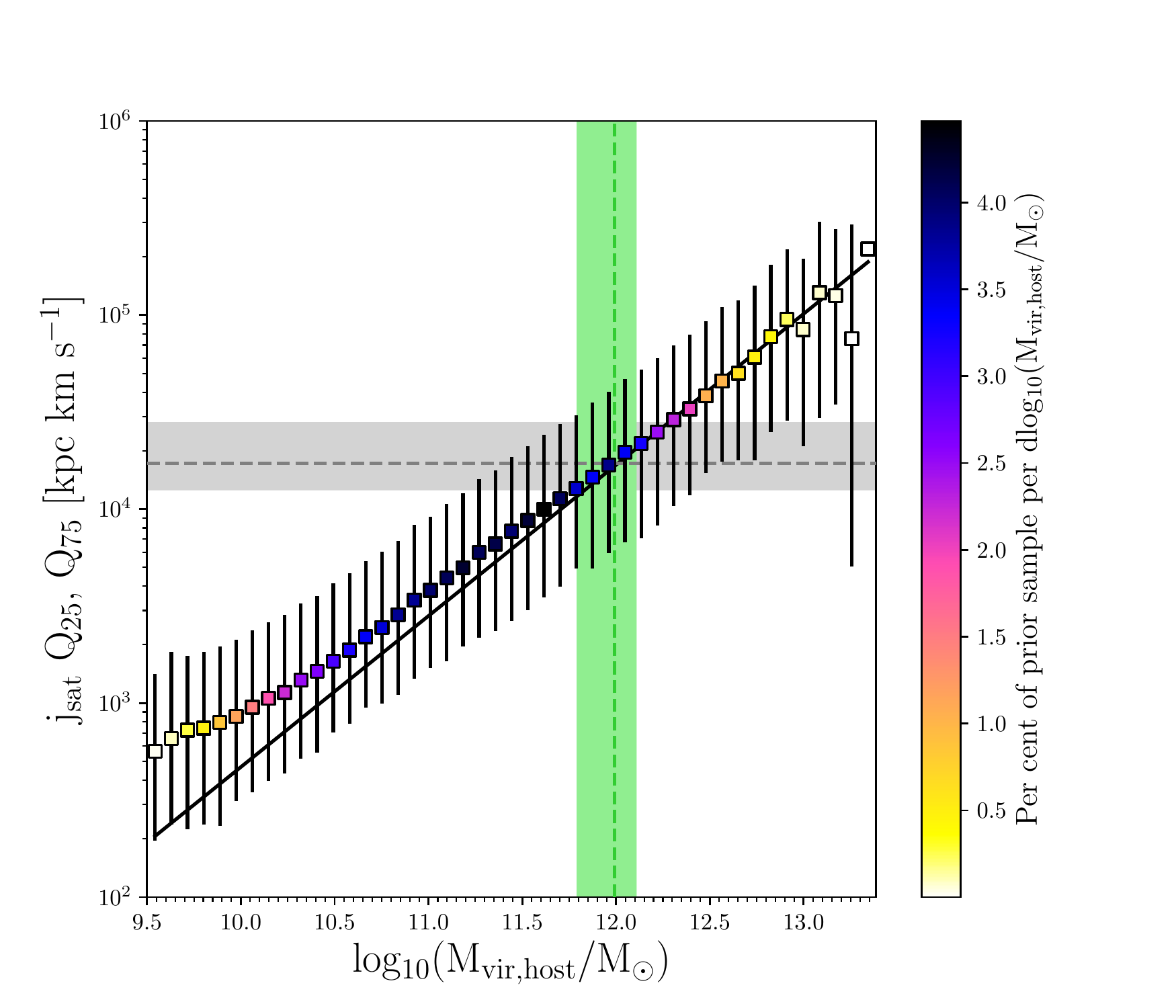}
\caption{Similar to Fig.~\ref{fig:mvirjrelation} except these subhalos reside in systems where a massive satellite analog (Section \ref{subsubsec:massivesats}) is not required. This sample is chosen from the final five snapshots of the Illustris-Dark halo catalogs and consists of 179,381 subhalos. The green dashed lines and shaded region now represent the ensemble posterior mean mass of the MW using eight satellites and the corresponding 68\% credible interval using the momentum likelihood and this new data sample ($\rm log_{10}(M_{vir}/M_{\odot})=11.99^{+0.12}_{-0.20}$). The black solid line is the line of best fit from Fig.~\ref{fig:mvirjrelation}.  The general relationship between host halo mass and specific orbital angular momenta holds, except at the low mass end. In low mass host halos, the angular momentum distribution is on average higher if massive satellite analogs are not included. Without a massive satellite analog, subhalos are most likely to be found in halos with masses ranging from log$\rm_{10}$(\Mvir/\Msun) $\approx$ 10.6-12.1. The agreement in MW mass estimates (green shaded region) between these results and Fig. \ref{fig:mvirjrelation} shows that the method is not strongly biased by the selection of the prior sample. 
\label{fig:mvirjrelationnoLMC}}
\end{center}
\end{figure}

\subsection{The Halo Mass-Specific Angular Momenta Distribution Relation in the Absence of a Massive Satellite}
We now explore whether the absence of a massive satellite analog changes the satellite specific angular momentum distribution as a function of host halo mass. To do so, we choose a new sample including all host halos in the final five snapshots of Illustris-Dark that pass the low mass satellite analog selection criteria (${\bf C'}$) given in Section \ref{subsubsec:lowmasssats} without requiring that a massive satellite analog (Section \ref{subsubsec:massivesats}) also resides in the halo. Thus, the host halos included in Prior 1 are strictly excluded from this sample. The new sample contains 179,381 low mass satellite analogs residing in 104,362 unique host halos.

Fig. \ref{fig:mvirjrelationnoLMC} shows the distribution of subhalo specific orbital angular momenta binned by halo mass for this alternate set of Illustris-Dark subhalos. These subhalos reside in halos with masses extending down to $\rm log_{10}(M_{vir}/M_{\odot})\approx 9.3$ due to the absence of a massive satellite analog, roughly two orders of magnitude lower than in Fig. \ref{fig:mvirjrelation}.} The gray shaded region is identical to that shown in Fig. \ref{fig:mvirjrelation} and represents the range of specific orbital angular momenta for eight low mass classical MW satellites. The green shaded region represents the posterior mean and 68\% credible interval included in the ensemble MW mass estimate from the momentum method calculated using this new sample (i.e. no massive satellite analogs). 
 
The black solid line is the line of best fit from Fig.~\ref{fig:mvirjrelation} and it indicates that the overall relationship between halo mass and specific orbital angular momentum distribution (i.e. the slope of the distribution) still exists. However, at very low host halo masses, the corresponding orbital angular momentum distribution begins to deviate from this trend such that the median specific orbital angular momentum is higher than if a massive satellite analog is included. This is important because it means that massive satellites (i.e. $\sim$ 10\% of their host's mass) may affect the orbital histories and kinematics of lower mass satellites, but it is harder to see explicitly in the higher host halo masses because a wide range of massive satellite analogs were included in the analysis. We conclude that there is a relationship between these two quantities only in host halos with masses $\rm log_{10}(M_{vir}/M_{\odot})> 11.5$ whether or not a massive satellite analog (i.e. LMC-mass companion) is present.

A few additional points of interest are also worth noting upon comparing Figs. \ref{fig:mvirjrelation} and \ref{fig:mvirjrelationnoLMC}. The abundance of low mass satellite analogs as a function of host halo mass (depicted by the color bars) is highest in halos with $\rm log_{10}(M_{vir}/M_{\odot})\approx 11.8-12.3$ in Fig. \ref{fig:mvirjrelation} when massive satellite analogs are present compared to a range of $\rm log_{10}$(\Mvir/\Msun) $\approx$ 10.6-12.1 in Fig. \ref{fig:mvirjrelationnoLMC} when they are not present. Such differences are expected from the hierarchical assembly of galaxy halos and are probably closely linked to the subhalo abundance functions for halos with and without a massive satellite companion (see Appendix \ref{sec:appendix1}). Despite these differences, the analysis still yields narrow MW mass ranges that are in agreement with each other (green shaded regions). This suggests there is no bias associated with the presence or absence of a massive satellite analog and that our prior selection criteria are justified. 

The existence of a strong correlation between halo mass and the distribution of orbital angular momenta for subhalos provides a generalized trend that can be applied to halos and their subsequent satellite populations. While the MW system is often considered common in a cosmological context, recent work by \citet{geha17} cautions against thinking about the MW as a {\em typical} halo in their study of observational MW analogs and their satellite populations' abundance and properties. Our notes on this trend are therefore widely applicable. 

\section{Conclusions}
\label{sec:conclusion}
Building on the Bayesian framework used to estimate the MW's virial mass from P17B, we provide a new methodology to combine the 6D phase space information of multiple low mass dwarf satellites. By doing so, we are able to estimate the mass of the MW using each of the following satellites individually and as an ensemble: Fornax, Sculptor, Carina, Draco, Leo I, Ursa Minor, Sextans, Leo II, Sagittarius dSph. This method is straightforwardly generalizable to include more dwarf satellites as new data is obtained. The main conclusions of this work are summarized below.

\begin{itemize}[leftmargin=*]
\item When the mass of the MW is inferred using each low mass satellite individually, we find a much larger scatter in the mass of the MW using the instantaneous method. This method yields a MW mass range of $0.6-2\times10^{12}$ \Msun across eight low mass satellites (Sagittarius dSph excluded), while the momentum method spans a much narrower range from $0.9-1.5\times10^{12}$ \Msun for eight satellites. The latter results are more constrained than the current values in the literature. 

\item By combining the posterior distributions associated with each satellite, we also report ensemble MW mass estimates. The combination of eight low mass satellites with the instantaneous method results in a MW mass of $\rm M_{vir, MW}= 1.19^{+0.19}_{-0.21}\times 10^{12}$ \Msun ($\rm log_{10}(M_{vir}/M_{\odot})=12.08^{+0.06}_{-0.09}$). Using the momentum method, the ensemble mass estimate is $\rm M_{vir, MW}= 0.96^{+0.29}_{-0.28}\times10^{12}$ ($ \rm log_{10}(M_{vir}/M_{\odot})=11.98^{+0.11}_{-0.15}$). If Sagittarius dSph is included from the latter estimate, it decreases to $\rm M_{vir, MW}= 0.85^{+0.22}_{-0.26}\times 10^{12}$ \Msun ($\rm log_{10}(M_{vir}/M_{\odot})=11.93^{+0.11}_{-0.16}$). These ensemble estimates are more precise than the masses inferred by any individual satellite and they narrow the mass range to less than a factor of two. 

\item Satellites with high specific orbital angular momentum like Leo II, as opposed to high speed satellites like Leo I or the LMC, now constrain the upper end of the MW's mass range. Satellites in the midst of disruption like the Sagittarius dSph have the lowest specific orbital angular momentum and therefore push the MW's mass to low values. As any one satellite alone can result in a biased MW mass estimate, satellite galaxies that are not undergoing disruption (evidenced by tidal tails or streams) are the most ideal candidates for this method. 

\item The 68\% credible interval of the ensemble MW mass resulting from the momentum method when the Sagittarius dSph is excluded ranges from $0.67-1.25 \times 10^{12}$ \Msun. This suggests Leo I could be bound to the MW if its true mass is in the upper half of this range (since \Mvir$=10^{12}$ \Msun corresponds to \Rvir $\approx$ 260 kpc). Our analysis only includes a fraction of the MW's satellite galaxy population, and while our MW mass estimates are not expected to change drastically as more satellites are included, these results are still preliminary.

\item We find there is a linear relationship between host halo mass and the distribution of specific orbital angular momentum for subhalos at a given halo mass. This trend is independent of the presence of massive satellite analogs and their orbital histories (i.e. first infall versus multiple orbits). The presence of a massive satellite analog does, however, change the halo mass range in which low mass subhalos are likely to reside. For example, when we require all halos to harbor a massive satellite analog, subhalos are most abundant in halos with masses $\rm log_{10}(M_{vir}/M_{\odot})\approx 11.8-12.3$. When massive satellites are not present, this range broadens and shifts down to $\rm log_{10}(M_{vir}/M_{\odot})\approx 10.6-12.1$. The median specific orbital angular momentum of satellites in lower mass host halos is higher if massive satellite analogs are not present, suggesting a massive satellite that is a high fraction of the host halo mass can affect the kinematics of the low mass satellite population. While this trend brackets the orbital angular momenta distribution expected at a given host halo mass, our tests on the conditional independence assumption shows that the individual angular momentum vectors of satellites around a shared host are independent of one another.

\item Upon comparing this trend to the distribution of specific orbital angular momenta for MW satellites with measured proper motions, we predict that ultra-faint MW dwarfs will have specific orbital angular momenta values between $5 \times 10^3 - 5 \times 10^4$ kpc \kms. Future proper motion measurements for MW dwarfs (HST-GO-14734, P.I. N. Kallivayalil) and specifically the ultra-faints residing at $\sim$ 100 kpc (HST-GO-14236, P.I. S.T. Sohn) will allow us to test this hypothesis. 
\end{itemize}

We have shown that combining a majority of the available satellite phase space information has already narrowed the plausible mass range for the MW, and eventually, the same can be done for M31 and other nearby galaxies in the era of JWST. As more satellite information becomes available, significant improvements to current cosmological simulations are crucial so a large prior sample of MW analogs, each hosting tens of satellites, can be selected. Together, a larger high precision data set for MW substructures and more advanced simulations will be powerful tools for converging on a precise and accurate MW mass.

\acknowledgments
EP is supported by the National Science Foundation through the Graduate Research Fellowship Program funded by Grant Award No. DGE-1746060. EP and GB are supported by NSF Grant AST-1714979. KM was supported in part by NSF grant AST-1516854. STS is supported by NASA through grant GO-12966. This research was also funded through a grant for HST program AR-12632. Support for AR-12632 and GO-12966 was provided by NASA through a grant from the Space Telescope Science Institute, which is operated by the Association of Universities for Research in Astronomy, Inc., under NASA contract NAS 5-26555. The Illustris simulations were run on the Odyssey cluster supported by the FAS Science Division Research Computing Group at Harvard University. We are also grateful to Roeland van der Marel, Laura Watkins, Mark Fardal, Dennis Zaritsky, Vicente Rodriguez-Gomez, Tom McClintock, Shea Garrison-Kimmel, and Hans-Walter Rix for useful discussions that have contributed to this paper.

\software{Example code developed for this project is available from \url{https://github.com/ekta1224/BayesToolsSatDynamics} under the MIT open-source software license. This research also utilized: \texttt{IPython} \citep{ipython}, \texttt{numpy} \citep{numpy}, \texttt{scipy} \citep{scipy}, and \texttt{matplotlib} \citep{matplotlib}. }

\bibliographystyle{yahapj}
\bibliography{myrefs}

\begin{thebibliography}{}
\providecommand\natexlab[1]{#1}
\providecommand\JournalTitle[1]{#1}

\bibitem[{{Belokurov} {et~al.}(2014){Belokurov}, {Koposov}, {Evans},
  {Pe{\~n}arrubia}, {Irwin}, {Smith}, {Lewis}, {Gieles}, {Wilkinson},
  {Gilmore}, {Olszewski}, \& {Niederste-Ostholt}}]{belokurov14}
{Belokurov}, V., {Koposov}, S.~E., {Evans}, N.~W., {et~al.} 2014,
  \href{http://dx.doi.org/10.1093/mnras/stt1862}{\JournalTitle{\mnras}, 437,
  116}

\bibitem[{{Besla} {et~al.}(2007){Besla}, {Kallivayalil}, {Hernquist},
  {Robertson}, {Cox}, {van der Marel}, \& {Alcock}}]{b07}
{Besla}, G., {Kallivayalil}, N., {Hernquist}, L., {et~al.} 2007,
  \href{http://dx.doi.org/10.1086/521385}{\JournalTitle{\apj}, 668, 949}

\bibitem[{{Besla} {et~al.}(2012){Besla}, {Kallivayalil}, {Hernquist}, {van der
  Marel}, {Cox}, \& {Kere{\v s}}}]{besla12}
---. 2012,
  \href{http://dx.doi.org/10.1111/j.1365-2966.2012.20466.x}{\JournalTitle{\mnras},
  421, 2109}

\bibitem[{Besla {et~al.}(2010)Besla, Kallivayalil, Hernquist, van~der Marel,
  Cox, \& Kereš}]{besla10}
Besla, G., Kallivayalil, N., Hernquist, L., {et~al.} 2010,
  \href{http://stacks.iop.org/2041-8205/721/i=2/a=L97}{\JournalTitle{The
  Astrophysical Journal Letters}, 721, L97}

\bibitem[{{Boylan-Kolchin} {et~al.}(2011{\natexlab{a}}){Boylan-Kolchin},
  {Besla}, \& {Hernquist}}]{bk11}
{Boylan-Kolchin}, M., {Besla}, G., \& {Hernquist}, L. 2011{\natexlab{a}},
  \href{http://dx.doi.org/10.1111/j.1365-2966.2011.18495.x}{\JournalTitle{\mnras},
  414, 1560}

\bibitem[{{Boylan-Kolchin} {et~al.}(2011{\natexlab{b}}){Boylan-Kolchin},
  {Bullock}, \& {Kaplinghat}}]{bk11b}
{Boylan-Kolchin}, M., {Bullock}, J.~S., \& {Kaplinghat}, M. 2011{\natexlab{b}},
  \href{http://dx.doi.org/10.1111/j.1745-3933.2011.01074.x}{\JournalTitle{\mnras},
  415, L40}

\bibitem[{{Boylan-Kolchin} {et~al.}(2012){Boylan-Kolchin}, {Bullock}, \&
  {Kaplinghat}}]{bk12}
---. 2012,
  \href{http://dx.doi.org/10.1111/j.1365-2966.2012.20695.x}{\JournalTitle{\mnras},
  422, 1203}

\bibitem[{{Boylan-Kolchin} {et~al.}(2013){Boylan-Kolchin}, {Bullock}, {Sohn},
  {Besla}, \& {van der Marel}}]{bk13}
{Boylan-Kolchin}, M., {Bullock}, J.~S., {Sohn}, S.~T., {Besla}, G., \& {van der
  Marel}, R.~P. 2013,
  \href{http://dx.doi.org/10.1088/0004-637X/768/2/140}{\JournalTitle{\apj},
  768, 140}

\bibitem[{{Bryan} \& {Norman}(1998)}]{brynorman98}
{Bryan}, G.~L., \& {Norman}, M.~L. 1998,
  \href{http://dx.doi.org/10.1086/305262}{\JournalTitle{\apj}, 495, 80}

\bibitem[{{Busha} {et~al.}(2011){Busha}, {Marshall}, {Wechsler}, {Klypin}, \&
  {Primack}}]{busha11}
{Busha}, M.~T., {Marshall}, P.~J., {Wechsler}, R.~H., {Klypin}, A., \&
  {Primack}, J. 2011,
  \href{http://dx.doi.org/10.1088/0004-637X/743/1/40}{\JournalTitle{\apj}, 743,
  40}

\bibitem[{{Casetti-Dinescu} {et~al.}(2018){Casetti-Dinescu}, {Girard}, \&
  {Schriefer}}]{casetti17}
{Casetti-Dinescu}, D.~I., {Girard}, T.~M., \& {Schriefer}, M. 2018,
  \href{http://dx.doi.org/10.1093/mnras/stx2645}{\JournalTitle{\mnras}, 473,
  4064}

\bibitem[{{Chua} {et~al.}(2016){Chua}, {Pillepich}, {Rodriguez-Gomez},
  {Vogelsberger}, {Bird}, \& {Hernquist}}]{chua16}
{Chua}, K.~T.~E., {Pillepich}, A., {Rodriguez-Gomez}, V., {et~al.} 2016,
  \JournalTitle{ArXiv e-prints},
  \href{http://arxiv.org/abs/1611.07991}{{\sffamily arXiv:1611.07991}}

\bibitem[{{Dolag} {et~al.}(2009){Dolag}, {Borgani}, {Murante}, \&
  {Springel}}]{dolag09}
{Dolag}, K., {Borgani}, S., {Murante}, G., \& {Springel}, V. 2009,
  \href{http://dx.doi.org/10.1111/j.1365-2966.2009.15034.x}{\JournalTitle{\mnras},
  399, 497}

\bibitem[{{D'Onghia} {et~al.}(2010){D'Onghia}, {Springel}, {Hernquist}, \&
  {Keres}}]{donghia10}
{D'Onghia}, E., {Springel}, V., {Hernquist}, L., \& {Keres}, D. 2010,
  \href{http://dx.doi.org/10.1088/0004-637X/709/2/1138}{\JournalTitle{\apj},
  709, 1138}

\bibitem[{{Drlica-Wagner} {et~al.}(2015){Drlica-Wagner}, {Bechtol}, {Rykoff},
  {Luque}, {Queiroz}, {Mao}, {Wechsler}, {Simon}, {Santiago}, {Yanny},
  {Balbinot}, {Dodelson}, {Fausti Neto}, {James}, {Li}, {Maia}, {Marshall},
  {Pieres}, {Stringer}, {Walker}, {Abbott}, {Abdalla}, {Allam},
  {Benoit-L{\'e}vy}, {Bernstein}, {Bertin}, {Brooks}, {Buckley-Geer}, {Burke},
  {Carnero Rosell}, {Carrasco Kind}, {Carretero}, {Crocce}, {da Costa},
  {Desai}, {Diehl}, {Dietrich}, {Doel}, {Eifler}, {Evrard}, {Finley},
  {Flaugher}, {Fosalba}, {Frieman}, {Gaztanaga}, {Gerdes}, {Gruen}, {Gruendl},
  {Gutierrez}, {Honscheid}, {Kuehn}, {Kuropatkin}, {Lahav}, {Martini},
  {Miquel}, {Nord}, {Ogando}, {Plazas}, {Reil}, {Roodman}, {Sako}, {Sanchez},
  {Scarpine}, {Schubnell}, {Sevilla-Noarbe}, {Smith}, {Soares-Santos},
  {Sobreira}, {Suchyta}, {Swanson}, {Tarle}, {Tucker}, {Vikram}, {Wester},
  {Zhang}, {Zuntz}, \& {DES Collaboration}}]{dwagner15}
{Drlica-Wagner}, A., {Bechtol}, K., {Rykoff}, E.~S., {et~al.} 2015,
  \href{http://dx.doi.org/10.1088/0004-637X/813/2/109}{\JournalTitle{\apj},
  813, 109}

\bibitem[{{Garrison-Kimmel} {et~al.}(2017){Garrison-Kimmel}, {Wetzel},
  {Bullock}, {Hopkins}, {Boylan-Kolchin}, {Faucher-Giguere}, {Keres},
  {Quataert}, {Sanderson}, {Graus}, \& {Kelley}}]{garrisonkimmel17}
{Garrison-Kimmel}, S., {Wetzel}, A.~R., {Bullock}, J.~S., {et~al.} 2017,
  \JournalTitle{ArXiv e-prints},
  \href{http://arxiv.org/abs/1701.03792}{{\sffamily arXiv:1701.03792}}

\bibitem[{{Geha} {et~al.}(2017){Geha}, {Wechsler}, {Mao}, {Tollerud}, {Weiner},
  {Bernstein}, {Hoyle}, {Marchi}, {Marshall}, {Munoz}, \& {Lu}}]{geha17}
{Geha}, M., {Wechsler}, R.~H., {Mao}, Y.-Y., {et~al.} 2017, \JournalTitle{ArXiv
  e-prints}, \href{http://arxiv.org/abs/1705.06743}{{\sffamily
  arXiv:1705.06743}}

\bibitem[{{Genel} {et~al.}(2014){Genel}, {Vogelsberger}, {Springel}, {Sijacki},
  {Nelson}, {Snyder}, {Rodriguez-Gomez}, {Torrey}, \& {Hernquist}}]{genel14}
{Genel}, S., {Vogelsberger}, M., {Springel}, V., {et~al.} 2014,
  \href{http://dx.doi.org/10.1093/mnras/stu1654}{\JournalTitle{\mnras}, 445,
  175}

\bibitem[{{Gibbons} {et~al.}(2014){Gibbons}, {Belokurov}, \&
  {Evans}}]{gibbons14}
{Gibbons}, S.~L.~J., {Belokurov}, V., \& {Evans}, N.~W. 2014,
  \href{http://dx.doi.org/10.1093/mnras/stu1986}{\JournalTitle{\mnras}, 445,
  3788}

\bibitem[{{Gonz{\'a}lez} {et~al.}(2013){Gonz{\'a}lez}, {Kravtsov}, \&
  {Gnedin}}]{gonzalez13}
{Gonz{\'a}lez}, R.~E., {Kravtsov}, A.~V., \& {Gnedin}, N.~Y. 2013,
  \href{http://dx.doi.org/10.1088/0004-637X/770/2/96}{\JournalTitle{\apj}, 770,
  96}

\bibitem[{{Hinshaw} {et~al.}(2013){Hinshaw}, {Larson}, {Komatsu}, {Spergel},
  {Bennett}, {Dunkley}, {Nolta}, {Halpern}, {Hill}, {Odegard}, {Page}, {Smith},
  {Weiland}, {Gold}, {Jarosik}, {Kogut}, {Limon}, {Meyer}, {Tucker}, {Wollack},
  \& {Wright}}]{hinshaw13}
{Hinshaw}, G., {Larson}, D., {Komatsu}, E., {et~al.} 2013,
  \href{http://dx.doi.org/10.1088/0067-0049/208/2/19}{\JournalTitle{\apjs},
  208, 19}

\bibitem[{Hunter(2007)}]{matplotlib}
Hunter, J.~D. 2007,
  \href{http://dx.doi.org/10.1109/MCSE.2007.55}{\JournalTitle{Computing in
  Science Engineering}, 9, 90}

\bibitem[{Jones {et~al.}(2001--)Jones, Oliphant, Peterson, {et~al.}}]{scipy}
Jones, E., Oliphant, T., Peterson, P., {et~al.} 2001--, {SciPy}: Open source
  scientific tools for {Python}, [Online; accessed 2017-02-02]

\bibitem[{{Kafle} {et~al.}(2012){Kafle}, {Sharma}, {Lewis}, \&
  {Bland-Hawthorn}}]{kafle12}
{Kafle}, P.~R., {Sharma}, S., {Lewis}, G.~F., \& {Bland-Hawthorn}, J. 2012,
  \href{http://dx.doi.org/10.1088/0004-637X/761/2/98}{\JournalTitle{\apj}, 761,
  98}

\bibitem[{{Kafle} {et~al.}(2014){Kafle}, {Sharma}, {Lewis}, \&
  {Bland-Hawthorn}}]{kafle14}
---. 2014,
  \href{http://dx.doi.org/10.1088/0004-637X/794/1/59}{\JournalTitle{\apj}, 794,
  59}

\bibitem[{{Kallivayalil} {et~al.}(2006{\natexlab{a}}){Kallivayalil}, {van der
  Marel}, \& {Alcock}}]{k06b}
{Kallivayalil}, N., {van der Marel}, R.~P., \& {Alcock}, C. 2006{\natexlab{a}},
  \href{http://dx.doi.org/10.1086/508014}{\JournalTitle{\apj}, 652, 1213}

\bibitem[{{Kallivayalil} {et~al.}(2006{\natexlab{b}}){Kallivayalil}, {van der
  Marel}, {Alcock}, {Axelrod}, {Cook}, {Drake}, \& {Geha}}]{k06a}
{Kallivayalil}, N., {van der Marel}, R.~P., {Alcock}, C., {et~al.}
  2006{\natexlab{b}},
  \href{http://dx.doi.org/10.1086/498972}{\JournalTitle{\apj}, 638, 772}

\bibitem[{{Kallivayalil} {et~al.}(2013){Kallivayalil}, {van der Marel},
  {Besla}, {Anderson}, \& {Alcock}}]{k13}
{Kallivayalil}, N., {van der Marel}, R.~P., {Besla}, G., {Anderson}, J., \&
  {Alcock}, C. 2013,
  \href{http://dx.doi.org/10.1088/0004-637X/764/2/161}{\JournalTitle{\apj},
  764, 161}

\bibitem[{{Kim} {et~al.}(1998){Kim}, {Staveley-Smith}, {Dopita}, {Freeman},
  {Sault}, {Kesteven}, \& {McConnell}}]{kim98}
{Kim}, S., {Staveley-Smith}, L., {Dopita}, M.~A., {et~al.} 1998,
  \href{http://dx.doi.org/10.1086/306030}{\JournalTitle{\apj}, 503, 674}

\bibitem[{{Li} \& {White}(2008)}]{liwhite08}
{Li}, Y.-S., \& {White}, S.~D.~M. 2008,
  \href{http://dx.doi.org/10.1111/j.1365-2966.2007.12748.x}{\JournalTitle{\mnras},
  384, 1459}

\bibitem[{{Li} {et~al.}(2017){Li}, {Jing}, {Qian}, {Yuan}, \& {Zhao}}]{li17}
{Li}, Z.-Z., {Jing}, Y.~P., {Qian}, Y.-Z., {Yuan}, Z., \& {Zhao}, D.-H. 2017,
  \href{http://dx.doi.org/10.3847/1538-4357/aa94c0}{\JournalTitle{\apj}, 850,
  116}

\bibitem[{{Massari} {et~al.}(2013){Massari}, {Bellini}, {Ferraro}, {van der
  Marel}, {Anderson}, {Dalessandro}, \& {Lanzoni}}]{massari13}
{Massari}, D., {Bellini}, A., {Ferraro}, F.~R., {et~al.} 2013,
  \href{http://dx.doi.org/10.1088/0004-637X/779/1/81}{\JournalTitle{\apj}, 779,
  81}

\bibitem[{{McConnachie}(2012)}]{mcconnachie12}
{McConnachie}, A.~W. 2012,
  \href{http://dx.doi.org/10.1088/0004-6256/144/1/4}{\JournalTitle{\aj}, 144,
  4}

\bibitem[{{Moster} {et~al.}(2013){Moster}, {Naab}, \& {White}}]{moster13}
{Moster}, B.~P., {Naab}, T., \& {White}, S.~D.~M. 2013,
  \href{http://dx.doi.org/10.1093/mnras/sts261}{\JournalTitle{\mnras}, 428,
  3121}

\bibitem[{{Nelson} {et~al.}(2015){Nelson}, {Pillepich}, {Genel},
  {Vogelsberger}, {Springel}, {Torrey}, {Rodriguez-Gomez}, {Sijacki}, {Snyder},
  {Griffen}, {Marinacci}, {Blecha}, {Sales}, {Xu}, \& {Hernquist}}]{nelson15}
{Nelson}, D., {Pillepich}, A., {Genel}, S., {et~al.} 2015,
  \href{http://dx.doi.org/10.1016/j.ascom.2015.09.003}{\JournalTitle{Astronomy
  and Computing}, 13, 12}

\bibitem[{{Newton} {et~al.}(2017){Newton}, {Cautun}, {Jenkins}, {Frenk}, \&
  {Helly}}]{newton17}
{Newton}, O., {Cautun}, M., {Jenkins}, A., {Frenk}, C.~S., \& {Helly}, J. 2017,
  \JournalTitle{ArXiv e-prints},
  \href{http://arxiv.org/abs/1708.04247}{{\sffamily arXiv:1708.04247}}

\bibitem[{{Pasetto} {et~al.}(2011){Pasetto}, {Grebel}, {Berczik}, {Chiosi}, \&
  {Spurzem}}]{pasetto11}
{Pasetto}, S., {Grebel}, E.~K., {Berczik}, P., {Chiosi}, C., \& {Spurzem}, R.
  2011,
  \href{http://dx.doi.org/10.1051/0004-6361/200913415}{\JournalTitle{\aap},
  525, A99}

\bibitem[{{Patel} {et~al.}(2017b){Patel}, {Besla}, \& {Mandel}}]{patel17b}
{Patel}, E., {Besla}, G., \& {Mandel}, K. 2017b,
  \href{http://dx.doi.org/10.1093/mnras/stx698}{\JournalTitle{\mnras}, 468,
  3428}

\bibitem[{{Patel} {et~al.}(2017a){Patel}, {Besla}, \& {Sohn}}]{patel17a}
{Patel}, E., {Besla}, G., \& {Sohn}, S.~T. 2017a,
  \href{http://dx.doi.org/10.1093/mnras/stw2616}{\JournalTitle{\mnras}, 464,
  3825}

\bibitem[{Perez \& Granger(2007)}]{ipython}
Perez, F., \& Granger, B.~E. 2007,
  \href{http://dx.doi.org/10.1109/MCSE.2007.53}{\JournalTitle{Computing in
  Science Engineering}, 9, 21}

\bibitem[{{Piatek} {et~al.}(2005){Piatek}, {Pryor}, {Bristow}, {Olszewski},
  {Harris}, {Mateo}, {Minniti}, \& {Tinney}}]{piatek05}
{Piatek}, S., {Pryor}, C., {Bristow}, P., {et~al.} 2005,
  \href{http://dx.doi.org/10.1086/430532}{\JournalTitle{\aj}, 130, 95}

\bibitem[{{Piatek} {et~al.}(2006){Piatek}, {Pryor}, {Bristow}, {Olszewski},
  {Harris}, {Mateo}, {Minniti}, \& {Tinney}}]{piatek06}
---. 2006, \href{http://dx.doi.org/10.1086/499526}{\JournalTitle{\aj}, 131,
  1445}

\bibitem[{{Piatek} {et~al.}(2007){Piatek}, {Pryor}, {Bristow}, {Olszewski},
  {Harris}, {Mateo}, {Minniti}, \& {Tinney}}]{piatek07}
---. 2007, \href{http://dx.doi.org/10.1086/510456}{\JournalTitle{\aj}, 133,
  818}

\bibitem[{{Piatek} {et~al.}(2008){Piatek}, {Pryor}, \& {Olszewski}}]{piatek08}
{Piatek}, S., {Pryor}, C., \& {Olszewski}, E.~W. 2008,
  \href{http://dx.doi.org/10.1088/0004-6256/135/3/1024}{\JournalTitle{\aj},
  135, 1024}

\bibitem[{{Piatek} {et~al.}(2016){Piatek}, {Pryor}, \& {Olszewski}}]{piatek16}
---. 2016,
  \href{http://dx.doi.org/10.3847/0004-6256/152/6/166}{\JournalTitle{\aj}, 152,
  166}

\bibitem[{{Piatek} {et~al.}(2003){Piatek}, {Pryor}, {Olszewski}, {Harris},
  {Mateo}, {Minniti}, \& {Tinney}}]{piatek03}
{Piatek}, S., {Pryor}, C., {Olszewski}, E.~W., {et~al.} 2003,
  \href{http://dx.doi.org/10.1086/378713}{\JournalTitle{\aj}, 126, 2346}

\bibitem[{{Piatek} {et~al.}(2002){Piatek}, {Pryor}, {Olszewski}, {Harris},
  {Mateo}, {Minniti}, {Monet}, {Morrison}, \& {Tinney}}]{piatek02}
---. 2002, \href{http://dx.doi.org/10.1086/344767}{\JournalTitle{\aj}, 124,
  3198}

\bibitem[{{Rodriguez-Gomez} {et~al.}(2015){Rodriguez-Gomez}, {Genel},
  {Vogelsberger}, {Sijacki}, {Pillepich}, {Sales}, {Torrey}, {Snyder},
  {Nelson}, {Springel}, {Ma}, \& {Hernquist}}]{rg15}
{Rodriguez-Gomez}, V., {Genel}, S., {Vogelsberger}, M., {et~al.} 2015,
  \href{http://dx.doi.org/10.1093/mnras/stv264}{\JournalTitle{\mnras}, 449, 49}

\bibitem[{{Sales} {et~al.}(2007){Sales}, {Navarro}, {Abadi}, \&
  {Steinmetz}}]{sales07}
{Sales}, L.~V., {Navarro}, J.~F., {Abadi}, M.~G., \& {Steinmetz}, M. 2007,
  \href{http://dx.doi.org/10.1111/j.1365-2966.2007.12026.x}{\JournalTitle{\mnras},
  379, 1475}

\bibitem[{{Scholz} \& {Irwin}(1994)}]{scholz94}
{Scholz}, R.-D., \& {Irwin}, M.~J. 1994, in IAU Symposium, Vol. 161, Astronomy
  from Wide-Field Imaging, ed. H.~T. {MacGillivray}, 535

\bibitem[{{Sohn} {et~al.}(2013){Sohn}, {Besla}, {van der Marel},
  {Boylan-Kolchin}, {Majewski}, \& {Bullock}}]{sohn13}
{Sohn}, S.~T., {Besla}, G., {van der Marel}, R.~P., {et~al.} 2013,
  \href{http://dx.doi.org/10.1088/0004-637X/768/2/139}{\JournalTitle{\apj},
  768, 139}

\bibitem[{{Sohn} {et~al.}(2017){Sohn}, {Patel}, {Besla}, {van der Marel},
  {Bullock}, {Strigari}, {van de Ven}, \& {Walker}}]{sohn17}
{Sohn}, S.~T., {Patel}, E., {Besla}, G., {et~al.} 2017, \JournalTitle{ArXiv
  e-prints}, \href{http://arxiv.org/abs/1707.02593}{{\sffamily
  arXiv:1707.02593}}

\bibitem[{{Sohn} {et~al.}(2015){Sohn}, {van der Marel}, {Carlin}, {Majewski},
  {Kallivayalil}, {Law}, {Anderson}, \& {Siegel}}]{sohn15}
{Sohn}, S.~T., {van der Marel}, R.~P., {Carlin}, J.~L., {et~al.} 2015,
  \href{http://dx.doi.org/10.1088/0004-637X/803/2/56}{\JournalTitle{\apj}, 803,
  56}

\bibitem[{{Springel} {et~al.}(2001){Springel}, {White}, {Tormen}, \&
  {Kauffmann}}]{springel01}
{Springel}, V., {White}, S.~D.~M., {Tormen}, G., \& {Kauffmann}, G. 2001,
  \href{http://dx.doi.org/10.1046/j.1365-8711.2001.04912.x}{\JournalTitle{\mnras},
  328, 726}

\bibitem[{{Springel} {et~al.}(2008){Springel}, {Wang}, {Vogelsberger},
  {Ludlow}, {Jenkins}, {Helmi}, {Navarro}, {Frenk}, \& {White}}]{springel08a}
{Springel}, V., {Wang}, J., {Vogelsberger}, M., {et~al.} 2008,
  \href{http://dx.doi.org/10.1111/j.1365-2966.2008.14066.x}{\JournalTitle{\mnras},
  391, 1685}

\bibitem[{{Tollerud} {et~al.}(2011){Tollerud}, {Boylan-Kolchin}, {Barton},
  {Bullock}, \& {Trinh}}]{tollerud11}
{Tollerud}, E.~J., {Boylan-Kolchin}, M., {Barton}, E.~J., {Bullock}, J.~S., \&
  {Trinh}, C.~Q. 2011,
  \href{http://dx.doi.org/10.1088/0004-637X/738/1/102}{\JournalTitle{\apj},
  738, 102}

\bibitem[{{van der Marel} \& {Sahlmann}(2016)}]{vdMsahlmann}
{van der Marel}, R.~P., \& {Sahlmann}, J. 2016,
  \href{http://dx.doi.org/10.3847/2041-8205/832/2/L23}{\JournalTitle{\apjl},
  832, L23}

\bibitem[{van~der Walt {et~al.}(2011)van~der Walt, Colbert, \&
  Varoquaux}]{numpy}
van~der Walt, S., Colbert, S.~C., \& Varoquaux, G. 2011,
  \href{http://dx.doi.org/10.1109/MCSE.2011.37}{\JournalTitle{Computing in
  Science Engineering}, 13, 22}

\bibitem[{{Vogelsberger} {et~al.}(2014{\natexlab{a}}){Vogelsberger}, {Genel},
  {Springel}, {Torrey}, {Sijacki}, {Xu}, {Snyder}, {Bird}, {Nelson}, \&
  {Hernquist}}]{vogelsberger14a}
{Vogelsberger}, M., {Genel}, S., {Springel}, V., {et~al.} 2014{\natexlab{a}},
  \href{http://dx.doi.org/10.1038/nature13316}{\JournalTitle{\nat}, 509, 177}

\bibitem[{{Vogelsberger} {et~al.}(2014{\natexlab{b}}){Vogelsberger}, {Genel},
  {Springel}, {Torrey}, {Sijacki}, {Xu}, {Snyder}, {Nelson}, \&
  {Hernquist}}]{vogelsberger14b}
---. 2014{\natexlab{b}},
  \href{http://dx.doi.org/10.1093/mnras/stu1536}{\JournalTitle{\mnras}, 444,
  1518}

\bibitem[{{Walker} {et~al.}(2008){Walker}, {Mateo}, \& {Olszewski}}]{walker08}
{Walker}, M.~G., {Mateo}, M., \& {Olszewski}, E.~W. 2008,
  \href{http://dx.doi.org/10.1086/595586}{\JournalTitle{\apjl}, 688, L75}

\bibitem[{{Wang} {et~al.}(2006){Wang}, {Li}, {Kauffmann}, \& {De
  Lucia}}]{wang06}
{Wang}, L., {Li}, C., {Kauffmann}, G., \& {De Lucia}, G. 2006,
  \href{http://dx.doi.org/10.1111/j.1365-2966.2006.10669.x}{\JournalTitle{\mnras},
  371, 537}

\bibitem[{{Wang} {et~al.}(2017){Wang}, {Han}, {Cole}, {Frenk}, \&
  {Sawala}}]{wang17}
{Wang}, W., {Han}, J., {Cole}, S., {Frenk}, C., \& {Sawala}, T. 2017,
  \href{http://dx.doi.org/10.1093/mnras/stx1334}{\JournalTitle{\mnras}, 470,
  2351}

\bibitem[{{Zaritsky} {et~al.}(1989){Zaritsky}, {Olszewski}, {Schommer},
  {Peterson}, \& {Aaronson}}]{zaritsky89}
{Zaritsky}, D., {Olszewski}, E.~W., {Schommer}, R.~A., {Peterson}, R.~C., \&
  {Aaronson}, M. 1989,
  \href{http://dx.doi.org/10.1086/167947}{\JournalTitle{\apj}, 345, 759}

\end{thebibliography}

\appendix

\section{The Halo Mass-Specific Angular Momenta Relation in the Presence of a Galactic Disk}
\label{sec:appendix2}

\begin{figure}
\begin{center}
\includegraphics[scale=0.58, trim = 0mm 0mm 0mm 0mm]{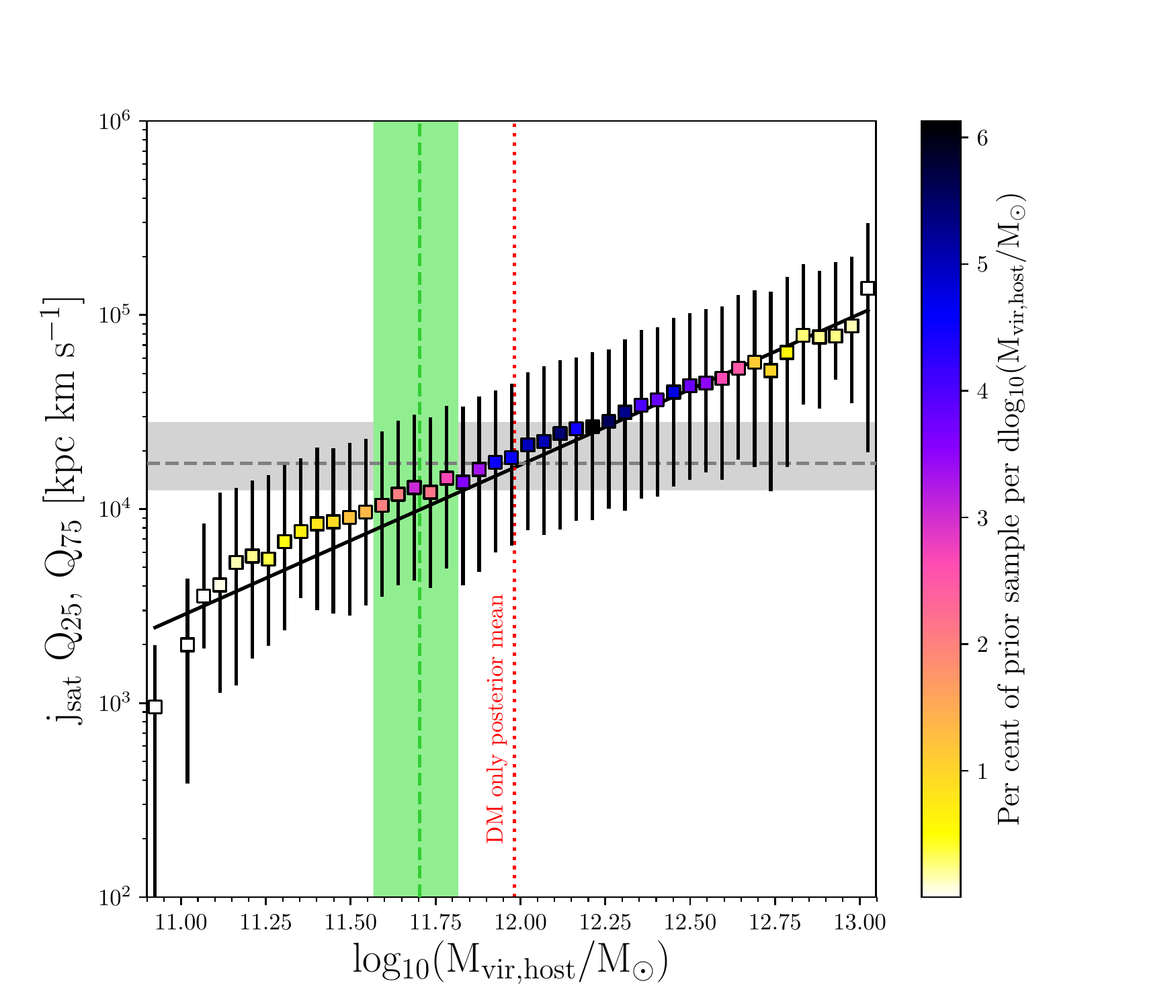}
\caption{ Similar to Fig. \ref{fig:mvirjrelation} but the data shown here are selected from the Illustris-1 simulation, which includes baryonic physics. This sample includes 59,156 subhalos residing in 18,858 unique host halos, just two-thirds of the subhalos found in Illustris-Dark. The black solid line is the line of best fit from Fig.~\ref{fig:mvirjrelation}. The green shaded region indicates the ensemble MW mass estimate ($\rm log_{10}(M_{vir}/M_{\odot}=11.70^{+0.12}_{-0.14}$) using the Illustris-1 data and the observed properties of 8 satellites. The Illustris-Dark results are indicated by the dotted red line. The overall trend between host halo mass and the distribution of specific orbital angular momenta is  in agreement with the dark-matter-only sample analysis beyond halo masses of $\rm log_{10}(M_{vir}/M_{\odot}) =11.75$. In host halos of lower mass, the kinematics of low mass subhalos may be more strongly affected by the co-evolution of dark matter and baryons, potentially causing the deviation from the line of best fit. The ensemble MW mass inferred with the Illustris-1 data is lower than that resulting from the dark-matter-only analysis, but they are consistent within $2\sigma$ of each other, suggesting that our technique is robust across simulations.
\label{fig:hydromvirjrelation}}
\end{center}
\end{figure} 

The presence of a galactic disk is known to cause a depletion of subhalos in the inner regions of MW-like halos \citep[e.g.][]{donghia10}. More specifically, recent work shows that subhalos on radial orbits are most susceptible to this phenomena \citep{garrisonkimmel17}. Here we use the Illustris-1 N-body+hydrodynamical simulation to examine whether there are any significant changes to the intrinsic relationship seen between host halo virial mass and the specific orbital angular momenta of subhalos discussed in Section \ref{sec:discussion} when baryons are considered. By comparing the results from the two versions of the Illustris simulation, we can also demonstrate whether the resulting MW mass estimates change significantly when both dark matter and baryons are taken into account.

To this end, a new sample of low mass satellite analogs are chosen from the final 20 snapshots of Illustris-1 that satisfy the same exact selection criteria described in Section \ref{subsec:prior}. This sample contains 59,156 subhalos residing in 18,858 unique host halos. Notice that one-third fewer subhalos satisfy our selection criteria in Illustris-1 compared to Illustris-Dark, indicating that there is a general depletion of low mass subhalos when disks and other hydrodynamic processes, such as feedback, are included. 

Fig. \ref{fig:hydromvirjrelation} shows the distribution of specific orbital angular momenta for subhalos in Illustris-1 binned by host halo virial mass. The gray shaded region is identical to the previous figures and represents the distribution of specific orbital angular momenta for the classical MW satellites. Fig. \ref{fig:hydromvirjrelation} is generally in good agreement with Figs. \ref{fig:mvirjrelation} and \ref{fig:mvirjrelationnoLMC}, as indicated by the consistency between the data points and the black solid line (the line of best fit from the dark-matter-only counterpart sample) for host halo masses above $\rm log_{10}(M_{vir}/M_{\odot}) =11.75$. The deviation from the line of best fit at host halo masses $\rm log_{10}(M_{vir}/M_{\odot}) < 11.75$ is likely due to the depletion of the lowest specific orbital angular momenta satellites and suggests there is a difference in subhalo kinematics between the simulations. The green shaded region indicates the ensemble MW mass estimate when the Illustris-1 data is used. Notice that the latter is lower than that predicted by the dark-matter-only subhalos, indicated by the red dotted line. 

We have also split the prior sample from Illustris-1 into subsets based on the time of their most recent major merger where this is defined as the last time there was a 1:4 stellar mass ratio collision such that the Illustris halo finding algorithm can no longer distinguish between two distinct halos. Hosts with major mergers greater than 6 Gyr ago make up 79\% of the sample and the halo mass-specific orbital angular momentum trend for this subset is in good agreement with Fig. \ref{fig:hydromvirjrelation}. The scatter associated with various formation histories is therefore implicitly included in our mass estimates even though our prior selection criteria do no explicitly require any specific formation history. Complementary work by \citet{li17} suggest that this scatter can contribute to up to 20\% uncertainty to the MW mass estimates.

We conclude that on average, the addition of baryonic physics and other hydrodynamical processes results in lower MW halo mass values. However, the posterior mean MW halo masses using each individual satellite and all satellites simultaneously are still within 2$\sigma$ of their counterpart MW posterior mean halo masses predicted by the dark-matter-only analysis. It appears that such discrepancy could arise from the depletion of low mass subhalos, which may consequently change the correlation between subhalo properties and their host halo properties. However, a new generation of cosmological simulations with revised feedback formulae and perhaps higher particle mass resolution may be able to address such concerns more confidently.

Since the subhalos considered here all have subhalo masses $\geq 10^9$ \Msun and most have a typical $v_{\rm max}=20-45$ \kms, their overall depletion is less drastic than expected. At this mass, subhalos are less susceptible to rapid tidal disruption by their hosts. Those subhalos that tend to experience significant depletion due to baryons are generally lower in mass and $v_{\rm max}$, such as in the simulation data studied by \citet{garrisonkimmel17}. If our method is extended to ultra-faint dwarf galaxies, these issues will need to be reconsidered.

\section{Subhalo Abundance Functions With and Without Massive Satellite Analogs}
\label{sec:appendix1}

\begin{figure}
\begin{center}
\includegraphics[scale=0.6, trim=8mm 0mm 10mm 5mm]{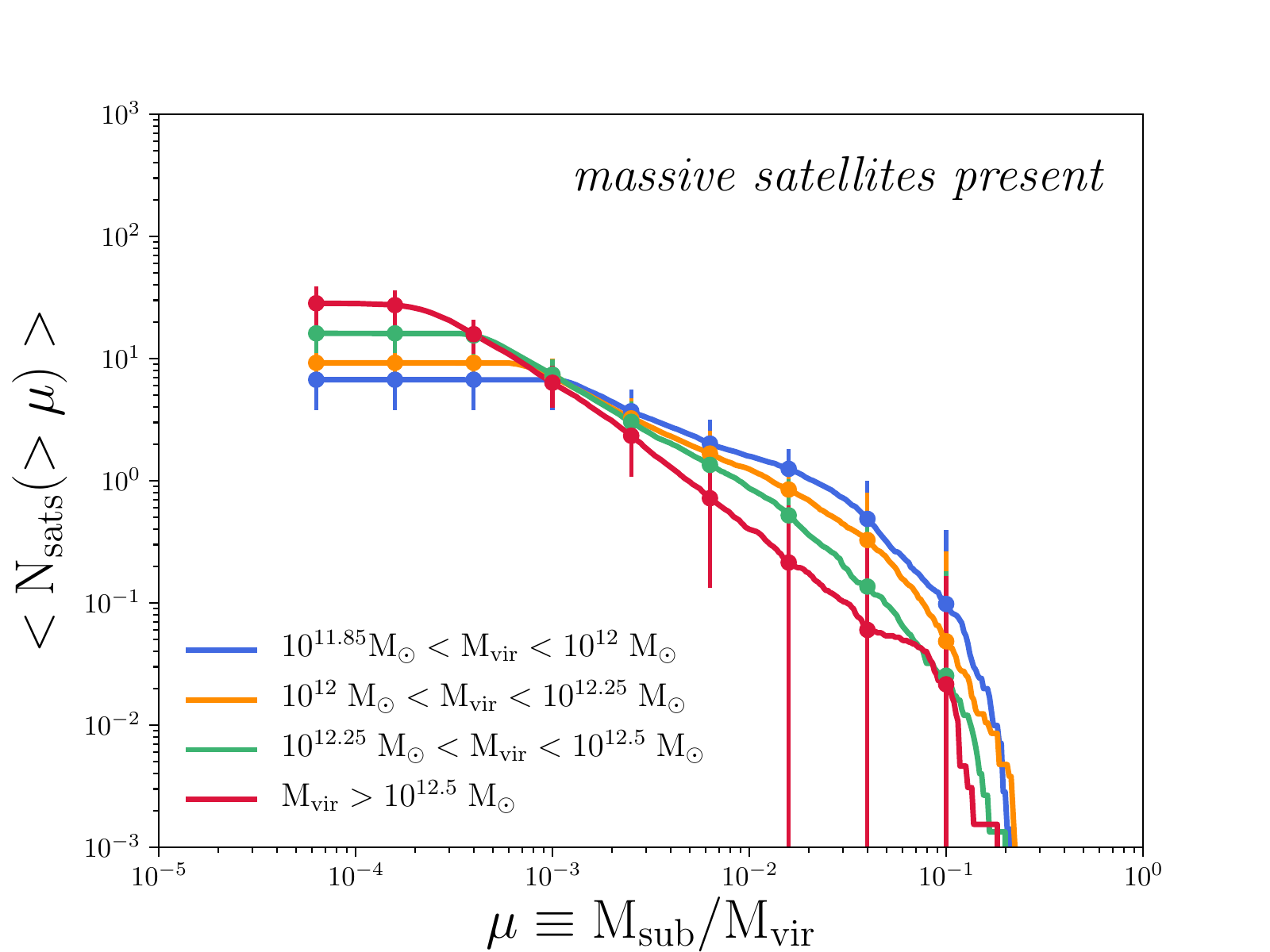}
\includegraphics[scale=0.6, trim=0mm 0mm 10mm 5mm]{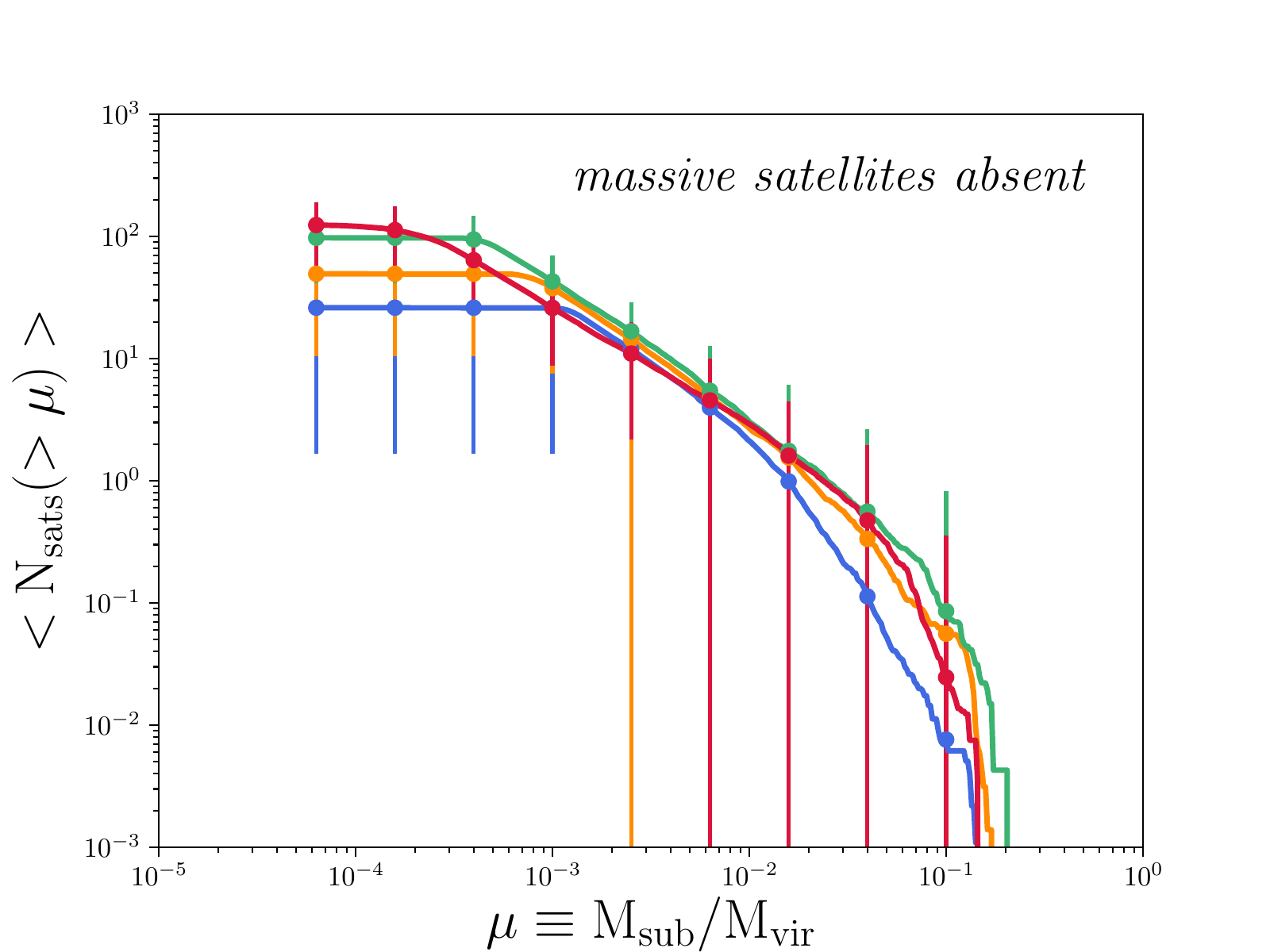}
\caption{{\em Left}: Subhalo abundance functions for the 3,153 unique host halos in Prior 2 with halo masses $< 10^{11.85}$ \Msun from the final 5 snapshots of Illustris-Dark. These hosts all have a massive satellite companion. All low mass subhalos around these hosts with a mass $\geq 10^9$ \Msun or 133 dark matter particles and that reside within the virial radius of their hosts are included. Abundances are plotted as a function of $\mu$ where $\mu$ is the ratio between subhalo mass and host virial mass ($M_{\rm sub}/M_{\rm vir}$). The data points with error bars indicate the standard deviation in each bin. Host halos with approximately 10 subhalos analogous to the classical MW satellites favor halo masses $10^{11.85} - 10^{12.25}$ \Msun, in agreement with our analysis results. {\em Right:} The subhalo abundance function for the 104,362 host halos that harbor low mass satellites analogs, but no massive satellite analog. This data is also taken from the final 5 snapshots of Illustris-Dark and all subhalos $\geq 10^9$ \Msun and within their host's virial radius are included. When massive satellite analogs are absent, on average, there are more low mass satellites about host halos with masses $ > 10^{11.85}$ \Msun. 
\label{fig:subabun}}
\end{center}
\end{figure}

In this analysis, all host halos with a massive satellite analog and one or more low mass satellite analogs were used to infer the halo mass of the MW (see Section \ref{sec:results}). The latter constraint, that there should be one or more low mass satellite analogs, rather than ten per host like the classical satellites around the MW, is necessary due to the current state of cosmological simulations and specifically their resolution limits. Here, we calculate the subhalo abundance functions around the host halos used for the MW halo mass calculations presented in Section \ref{sec:results} to demonstrate that on average, there are 10-20 subhalos with masses $\geq 10^9$ in the vicinity of their host's center of mass. However, the general properties of these subhalos, such as their $v_{\rm max}$ and current distances from their respective hosts, are not all representative of the classical MW satellites and were therefore omitted from Prior 2. In the left panel of Fig. \ref{fig:subabun}, we show the cumulative subhalo abundance function for all host halos in Prior 2 that were selected from the final 5 Illustris-Dark snapshots. These halos host exactly one massive satellite analog and at least one or more low mass satellite analogs (see Sections \ref{subsubsec:massivesats} and \ref{subsubsec:lowmasssats} for definitions). 

In Section \ref{sec:discussion}, we note that the trend between host halo mass and the distribution of satellite specific orbital angular momenta in the presence and absence of a massive satellite analog differs at host halo masses $< 10^{11.5}$ \Msun. Below this host halo mass, the median orbital angular momentum is higher when massive satellites are not present. However, inferred MW masses resulting from the ensemble of classical satellites are still in good agreement with each other, independent of this difference. Here, we examine the cumulative subhalo abundance for host halos that do not have a massive satellite analog since 33\% of $\sim 10^{12}$ \Msun halos typically host an LMC mass companion \citep{patel17a}. 

To create the abundance functions, all subhalos belonging to host halos in Prior 2, regardless of whether they satisfy our selection criteria in Sections \ref{subsubsec:massivesats} and \ref{subsubsec:lowmasssats}, with a subhalo mass $\geq 10^9$ \Msun and that reside within their host's virial radius are considered. All included subhalos are not necessarily members of the Prior 2 sample, but this sample encompasses all subhalos above our imposed $\sim 133$ dark matter particle resolution limit. The abundances are plotted as a function of $\mu \equiv M_{\rm sub}/M_{\rm vir}$, or the ratio between subhalo mass and host virial mass. The error bars on the data points indicate the standard deviations in each $\mu$ bin. Host halos less massive than $10^{11.85}$ \Msun are excluded, as this is beyond the range of MW masses in the literature.

The curves in the left panel of Fig. \ref{fig:subabun} all exhibit an increase near $\mu =10^{-2}$ due to the strict one massive satellite criterion (see Section \ref{subsubsec:massivesats}). This is most obvious in the lowest host halo masses (blue curve) where the massive satellites are a significant fraction of their host's mass. The highest mass host halos ($> 10^{12.5}$ \Msun, red line) are expected to host the greatest number of subhalos, but the strict one massive satellite criterion causes this trend to drop off more quickly around $\mu=10^{-3}$. Notice that at $\mu=10^{-3}$, all host halos harbor 8-10 subhalos on average. While the results presented in Section \ref{sec:results} use halos that have a massive satellite and one or more low mass satellite analogs, the abundance of subhalos around hosts in Prior 2 are generally in agreement with the observed MW satellite population.

In the right panel of Fig. \ref{fig:subabun}, the cumulative abundance functions are shown for all host halos that do not host a massive satellite analog. Notice there is no longer a peak near $\mu=10^{-2}$ as the requirement for exactly one massive satellite analog is eliminated. Overall, the shape of the abundance functions changes most significantly at $\mu > 10^{-3}$ and the abundance of subhalos is higher when a massive satellite analog is absent. At $\mu = 10^{-3}$, hosts tend to have 20-40 subhalos, compared to only 8-10 subhalos when massive satellites are present. The exact cause for the higher subhalo abundance in host halos $> 10^{11.85}$ \Msun is unknown and will be explored in future work. While the overall subhalo abundances differ for these host halo samples, the distribution of satellite specific orbital angular momentum for specifically the low mass satellite analogs in each sample generally remains the same and therefore yields similar MW masses.

Cumulative subhalo abundance functions for the Illustris simulations have been presented by \citet{chua16}, where they have not imposed any selection criteria on their host halos as we have implemented here. Their samples are chosen from the Illustris-1 simulation, which includes baryons and hydrodynamical processes, so the abundance of subhalos with masses $\sim 10^9$ \Msun is immediately lower (see Appendix \ref{sec:appendix2}). Using the sample chosen from Illustris-1, they find equivalent matches for each Illustris-1 halo in the Illustris-Dark simulation and formulate their cumulative abundances based on those halo populations. Our Fig. \ref{fig:subabun} data is chosen directly from the Illustris-Dark simulation and does not suffer these depletion effects, leading to higher subhalo abundances overall.

\end{document}